# Multicast and Unicast Superposition Transmission in MIMO OFDMA Systems with Statistical CSIT

Yong Jin Daniel Kim and David Vargas

*Abstract*—We consider a downlink multicast and unicast superposition transmission in muti-layer Multiple-Input Multiple-Output (MIMO) Orthogonal Frequency Division Multiple Access (OFDMA) systems when only the statistical channel state information is available at the transmitter (CSIT). Multiple users can be scheduled by using the time/frequency resources in OFDMA, while for each scheduled user MIMO spatial multiplexing is used to transmit multiple information layers, i.e., single user (SU)-MIMO. The users only need to feedback to the base-station the rank-indicator and the long-term average channel signal-to-noise ratio (SNR), i.e., statistical CSIT, to indicate a suitable number of transmission layers, a suitable modulation and coding scheme and allow the base-station to perform user scheduling. This approach is especially relevant for the delivery of common (e.g., popular live event) and independent (e.g., user personalized) content to a high number of users in deployments in the lower frequency bands operating in Frequency-Division-Duplex (FDD) mode, e.g., sub-1 GHz. We show that the optimal resource allocation that maximizes the ergodic sum-rate involves greedy user selection per OFDM subchannel and superposition transmission of one multicast signal across all subchannels and single unicast signal per subchannel. The optimal resource allocation dynamically chooses between sending unicast only, multicast only, or their superposition in each subchannel based on the disparity between the best and the worst user channel conditions and minimum multicast rate target. Degree-of-freedom (DoF) analysis shows that while the lack of instantaneous CSI limits DoF of unicast messages to the minimum number of transmit antennas and receiver antennas, the multicast message obtains full DoF that increases linearly with the number of users. We present resource allocation algorithms consisting of user selection and power allocation between multicast and unicast signals in each OFDM subchannel. System level simulations in 5G rural macro-cell scenarios show overall network throughput gains in realistic network environments by superposition transmission of multicast and unicast signals.

## I. Introduction

AS the consumption of media content (e.g., TV & radio) increasingly moves to the Internet, the efficient distribution of live and on-demand media experiences over mobile networks becomes important to service media companies. Users now enjoy popular live media streaming events (e.g., sports) combined with the engagement in social media networks while media experiences are also evolving to become increasingly personal, interactive, and immersive. For example, *Flexible Media* [1] allows the content of programmes to change according to the requirements of individual audience members (e.g., additional camera angles of a live sport event or changing the presenter for hearing impaired viewers).

Modern mobile broadband technologies have the capability to multiplex into the same system unicast signals (to convey independent information to each user, e.g., personalization) and broadcast/multicast signals (to efficiently deliver common information at the same time to multiple users, e.g., large live event). The 3GPP (3rd Generation Partnership Project) specified Multimedia Broadcast Multicast Services (MBMS) in the third and fourth generation mobile broadband technical specifications (3G and 4G) [2], and the Release-17 introduced Multicast Broadcast Services (MBS) into the 5G system [3], [4]. In 4G and 5G, broadcast/multicast and unicast signals are multiplexed using orthogonal time/frequency resources [5], [6] and scheduling between multicast and unicast has been considered in the literature [7].

Multiple-Input Multiple-Output (MIMO) is a key technology to provide high spectral efficiency in mobile systems and is an integral part of the 5G New Radio (NR) physical layer [8]. To increase the system's throughput compared to a single-input single-output (SISO) antenna systems, 4G and 5G specifications support single-user MIMO (SU-MIMO) and multi-user MIMO (MU-MIMO) transmission techniques [8], [9]. SU-MIMO spatially multiplexes multiple information streams (layers) within a given time/frequency resource to a single user where the multi-layer interference is managed at the receiver by multi-antenna receive processing. With SU-MIMO, multiple users can be scheduled for transmission in *orthogonal* time/frequency resources, where each user can receive multiple MIMO layers. MU-MIMO spatially multiplexes information streams of multiple users within the *same* time/frequency resource. The multi-stream interference in MU-MIMO is usually controlled at the transmitter by multi-antenna precoding, which requires knowledge of the channel state information at the transmitter (CSIT), and the receivers decode their desired signals from the corresponding spatial (or beamforming) directions while treating all other signals as noise. Assuming a transmitter sending unicast signals with $n_T$ antennas and users with $n_R$ antennas, the sum Degree of

Y. J. D. Kim is with the Department of Electrical and Computer Engineering, Rose-Hulman Institute of Technology, Terre Haute, IN 47803 USA (e-mail: kim2@rose-hulman.edu).

D. Vargas is with the BBC Research and Development, The Lighthouse, White City Place, 201 Wood Lane, London, W12 7TQ, U.K., (email: david.vargas@bbc.co.uk).

Freedom (DoF)[1] of SU-MIMO is $\min(n_T, n_R)$, while the sum DoF of MU-MIMO in the $K$-user scenario is $\min(n_T, Kn_R)$. This potential spatial multiplexing gain scaling with the number of users $K$ has attracted a lot of interest from academia and industry, and massive MIMO employing a large number of transmit antennas performs close to the capacity region of the MU-MIMO Gaussian broadcast channel with simple transmit linear processing [10]-[12].

As examples of large antenna arrays, field measurements at 3.5 and 6 GHz with 32 to 256 antenna elements are provided in [13]. At these frequencies, a large number of antennas can be integrated into an antenna system with a practical form factor size. However, at lower frequency bands, such as the sub-1 GHz with favorable propagation characteristics and covering large areas (e.g., rural), the number of antennas that can be integrated into an antenna system is significantly reduced. For example, at 700 MHz carrier frequency, a 4 by 8 antenna panel with antenna elements separated by $0.5\lambda$ (where $\lambda$ is the carrier wavelength) has a size of ~0.86m vertically by ~1.7m horizontally, which is not practical in most deployment scenarios [14], [15]. On the other hand, trials in [16] demonstrate significant performance gains with SU-MIMO in the 700 MHz band, and more recently tests in the same band in China demonstrated 4×4 MIMO downlink transmissions based on the 5G standards [17].

The performance gain of MU-MIMO with linear multi-antenna precoding is more limited in scenarios where the number of users served by the transmitter is similar or larger than the number of transmit antennas (i.e., in overloaded systems) [18]. At the sub-1 GHz band with a limited number of transmit antennas, the overloaded regime can be easily reached when serving a relatively small number of users and can be further accentuated when serving popular content to a large population of users in the cell. Additionally, most of the operating bands defined by 3GPP in NR for the sub-1 GHz spectrum use the Frequency Division Duplex mode (FDD) [8]. In FDD systems, users need to explicitly feedback the CSI required for multi-antenna precoding to the base-station, since uplink-downlink channel reciprocity cannot be exploited. 5G NR specifications support CSI reporting based on pre-defined codebooks of precoding vectors/matrices for both SU-MIMO (low-resolution Type-I codebook) and MU-MIMO (high-resolution Type-II codebook) [9], [19]. Although the time domain channel variation due to Doppler effect reduces with the carrier frequency, the associated feedback overhead and complexity at the user device for Type-II codebooks can still be a significant limiting factor for MU-MIMO at lower frequencies, users with high mobility and/or a high number of served users [9], [11].

Due to the potential limited MU-MIMO gain at the sub-1GHz band and the potential excessive overhead of high-resolution Type-II codebooks in FDD systems with high number of users and/or mobile users [9], it is important to study SU-MIMO systems with statistical CSIT where users only need to feedback the rank-indicator and the long-term average channel SNR to perform resource scheduling. Low-resolution Type-I codebook may also be considered for SU-MIMO, but this is left as a future work.

*A. Related work and contributions of this paper*

In this paper we consider a downlink multicast and unicast superposition transmission with muti-layer MIMO in Orthogonal Frequency Division Multiple Access (OFDMA) systems with statistical CSIT[2]. Multiple users can be scheduled in the system by using the time/frequency resources in OFDMA, while for each scheduled user MIMO spatial multiplexing is used to transmit multiple information layers (SU-MIMO). The users only need to feedback to the base-station the rank-indicator and the long-term average channel SNR (i.e., statistical CSIT) to indicate a suitable number of transmission layers, a suitable modulation and coding scheme (MCS) and allow the base-station to perform user scheduling. This approach is especially relevant for the delivery of common (e.g., popular live event) and independent (e.g., personalization) contents to a large number of users in the lower frequency bands operating in FDD mode, e.g., sub-1 GHz.

Inspired by the seminal work on superposition coding (SC) [20], [21], power-domain non-orthogonal multiplexing (NOM), also known as non-orthogonal-multiple-access (NOMA) or layered-division-multiplexing (LDM), is an alternative approach for joint multicast and unicast transmissions [22]-[35]. In superposition transmission, multicast and unicast signals with different power levels are superposed in the same time/frequency resources, where the power levels of each signal are chosen so that receivers can use successive interference cancellation (SIC) techniques to sequentially decode and cancel signals while treating previously un-decoded signals as noise. Under this framework, efficient scheduling of mixed multicast and unicast signals has been considered in the literature [22].

[23]-[31] consider MU-MIMO scenarios with perfect and imperfect CSIT to design precoder/beamformers with superposition transmission of multicast/broadcast and unicast under different constraints and applications: secrecy aspects [24], multi-cell scenarios with backhaul constraints [26] and imperfect CSIT [27], energy efficiency constraints [28], fronthaul compression and precoding in cloud radio networks [29], and simultaneous wireless information and power transfer [30]. A different approach based on Rate-Splitting (RS) is proposed in [31], in which dynamically chosen proportions of the unicast messages are combined into the multicast message so that they too can be decoded by all users. RS allows the reduction of the interference seen by each user and improves the performance compared to linear precoding or NOMA. Although the precoding design in RS also requires CSIT, RS is

---

[1] We say that the channel has $n$ DoF if the sum capacity can be written as $n\log(SNR) + o(\log(SNR))$. The DoF shows how many independent streams can be sent in the high SNR regime. A formal definition is given in section III.C.

[2] We consider the scenario where all the users have uplink capabilities and are connected to a base station.

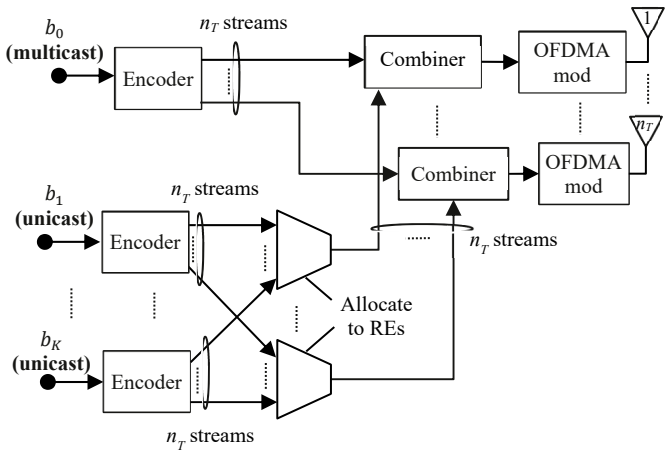

Fig. 1. Transmitter architecture of the multicast and unicast superposition transmission in MIMO OFDMA system.

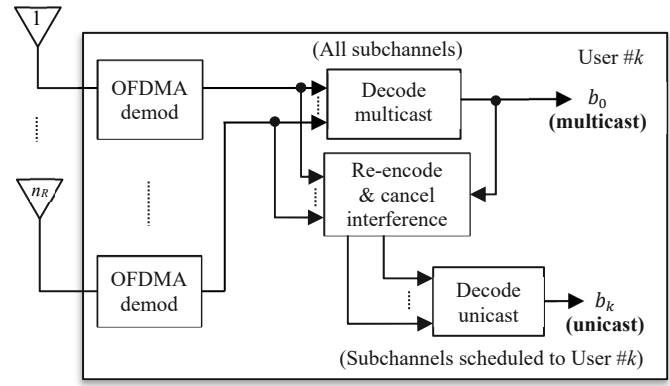

Fig. 2. Receiver architecture of user #$k$, which recovers both the multicast message that is sent to all users and the $k^{th}$ unicast message that is intended for the $k^{th}$ user only through successive interference cancellation.

more robust against CSIT inaccuracies than conventional linear precoding in MU-MIMO systems as demonstrated in [31].

With only statistical CSIT, spatial separation of users using precoding/beamforming is generally inaccessible (unless the channel is non-isotropic [36]), and user separation must be accomplished over remaining code/time/frequency/power radio resources. In [37], the ergodic capacity maximization problem in Rayleigh MIMO NOMA systems is considered, and the optimal input covariances as well as the optimal power allocations are found. The scope of [37], however, is limited to two users and a single time/frequency channel. For more than two users, weighted sum-rate and energy efficiency optimization problems in massive MIMO systems are considered in [38] and [39], respectively, but again a single channel is considered and multiple users are assumed to be spatially separable by leveraging the considered channel correlation model and properties of the massive MIMO systems. [32]-[35], [40] consider the application of superposition transmission of broadcast/multicast and unicast signals in cellular systems. [33], [34] calculates the rate region for a single broadcast signal (with Single-Input Multiple-Output (SIMO)) and a single unicast user (with multi-layer MIMO) in one (sub)channel. However, these works do not consider the $K$-user rate regions in OFDMA subchannels. Our prior work [35] on broadcast/multicast unicast superposition transmission (BMUST) in OFDMA systems only considers single antenna systems. Moreover, there has not yet been a theoretical analysis on how far the superposition of a multicast signal and a single unicast signal performs from the $K$-user downlink multiuser MIMO channel capacity with statistical CSIT.

The main contributions of this work are as follows:

- *To show that the optimal resource allocation that maximizes the ergodic sum rate involves greedy user selection per OFDM subchannel and superposition transmission of a multicast signal across all subchannels and a single unicast signal per subchannel.* The considered resource allocation problem is subject to transmit power constraint and minimum multicast rate target. The structure of optimal input covariances for both messages are also found. This work extends our theoretical results for BMUST in [35] for single antenna channels with full or statistical CSIT to multiuser MIMO channels with statistical CSIT for multicast and unicast superposition transmission.

The main implication of our work is that superposition beyond one multicast signal and one unicast signal is unnecessary in achieving the maximum sum-rate even with multiple users and multiple unicast messages. The delivery of more than one unicast signal may be accomplished by resource allocation over time/frequency resources, which motivates a study of an efficient resource allocation algorithm over OFDM subchannels.

- *To explicitly characterize the condition when sending unicast only, multicast only, or their superposition is optimal.* The choice depends on the disparity between the best user and worst user channel conditions and minimum multicast rate target.
- *To analyze the available DoF and show that the DoF of unicast is $min(n_T, n_R)$ and the DoF of multicast is $K \cdot min(n_T, n_R)$ where K is number of users and $n_T$ and $n_R$ are number of transmit and receive antennas, respectively.*
- *To present a resource allocation algorithm that consists of user selection and power allocation between multicast and unicast signals in each OFDM subchannel.* A simple near-optimal power allocation algorithm based on surrogate modeling of the ergodic rate function is proposed. Further simplification is obtained if only the power splits between the multicast and the unicast are optimized.
- *To present system level simulation results of the proposed resource allocation algorithms in 5G Rural Macro (RMa) scenarios.* These results reveal that a superposition transmission can provide significant performance improvements over various competing technologies including orthogonal multiplexing (OM), unicast-only and multicast-only transmissions in realistic environments.

The rest of the paper is organized as follows. Section II presents the considered channel model and the achievable rate region of the considered channel. The structure of the optimal transmission modes is derived in section III along with the DoF analysis. Section IV deals with efficient resource allocation over OFDM resources. System-level simulation results are provided in section V. Conclusions are given in section VI.

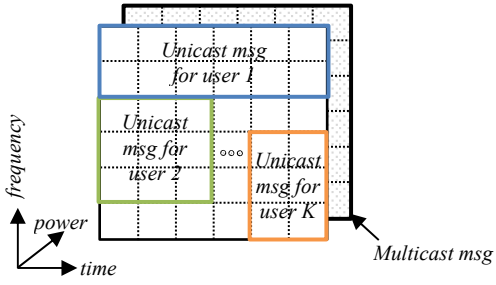

Fig. 3. Example time, frequency, and power allocation per transmit antenna. Splitting of the OFDM resources into subchannels consisting of arbitrary groupings of REs and allocation of the unicast messages to each subchannel. Multicast message is coded across the entire OFDM resources and superposition coded on top of the unicast signal.

*Notations:* $\mathbf{I}$ denotes identity matrix, $(\cdot)^\dagger$ denotes Hermitian of a matrix, $\det(\cdot)$ denotes matrix determinant, and $tr(\cdot)$ denotes trace of a matrix. Scalar matrices refer to diagonal matrices with equal values on the diagonal. $\min_{k \in \{1,2,...,K\}} \{R_k\}$ denotes $\min\{R_1, R_2, \ldots, R_K\}$ and $[x]^+$ denotes $\max\{x,0\}$. All logarithms have base 2.

## II. SYSTEM MODEL AND ACHIEVABLE RATE REGION

### A. Channel model and assumptions

A Rayleigh MIMO downlink channel with a transmitter with $n_T$ antennas and $K$ receivers, each with $n_R$ antennas, is considered. We consider a joint transmission of $K$ unicast messages and a single multicast message over OFDM resources. Fig. 1 and Fig. 2 show the reference transmit and receive architectures, respectively, for the multicast and unicast superposition transmission. The smallest time/frequency unit in OFDM is referred to as a resource element (RE). In 5G NR, one RE consists of one subcarrier in the frequency domain and one OFDM symbol in the time domain [8]. Without loss of generality, we divide the OFDM resources into $M$ "parallel" subchannels, each consisting of $N^{(1)}, \ldots, N^{(M)}$ REs, respectively, so that $\sum_{i=1}^{M} N^{(i)} = N$. This division of OFDM resources into $M$ subchannels is illustrated in Fig. 3. Messages are allocated to each subchannel based on a resource allocation algorithm to be discussed later. We allow the allocation of more than one message into one subchannel using non-orthogonal multiplexing.

Let $\mathcal{M} = \{1,2,...,M\}$ and $\mathcal{K} = \{1,2,...,K\}$. In subchannel $i \in \mathcal{M}$ at time $n$, user $k \in \mathcal{K}$ receives the $n_R \times 1$ received signal vector

$$\mathbf{y}_k^{(i)}[n] = \mathbf{H}_k^{(i)}[n]\mathbf{x}^{(i)}[n] + \mathbf{z}_k^{(i)}[n], \quad (1)$$

where $\mathbf{x}^{(i)}[n] \in \mathbb{C}^{n_T \times 1}$ is the symbol vector transmitted in subchannel $i$, $\mathbf{H}_k^{(i)}[n] \in \mathbb{C}^{n_R \times n_T}$ is the channel matrix of user $k$, and $\mathbf{z}_k^{(i)}[n]$ is normalized additive white Gaussian noise (AWGN) vector with unit variance. Assuming a rich scattering environment and memoryless channel, elements of the channel matrix $\mathbf{H}_k^{(i)}[n]$ are modeled as independent and identically distributed as $\mathcal{CN}\left(0, \sigma_{\mathbf{H}_k^{(i)}}^2\right)$ and independent in $n$. Different user channels $\mathbf{H}_k^{(i)}$ and $\mathbf{H}_r^{(i)}$ for $k \neq r$ may be correlated arbitrarily with one another. The effective long-term average channel SNR of user $k$ in subchannel $i$ is defined as

$$\sigma_{i,k}^2 \triangleq \sigma_{\mathbf{H}_k^{(i)}}^2 / \eta_i, \quad (2)$$

where $\eta_i \triangleq \frac{N^{(i)}}{N}$. $\sigma_{i,k}^2$ is simply referred to as *channel SNR* throughout this paper. Note that the noise power (which is unity) is scaled by $\eta_i$ to account for the $i^{\text{th}}$ subchannel consisting of $N^{(i)}$ out of $N$ REs. It is worth noting that same user can have different channel SNRs in different subchannels. We assume that the transmitter knows these channel SNRs (i.e., statistical CSIT), but not the instantaneous channel realizations (i.e., phases and amplitudes). Each receiver is assumed to have full CSI of its own channel $\mathbf{H}_k^{(i)}$ only for all $i$, through the standard channel estimation techniques using transmitted reference signal or pilot.

The Rayleigh MIMO channel with statistical CSIT falls into the class of stochastically degraded broadcast channels for which the superposition coding (SC) and the successive interference cancellation (SIC) are known to be optimal [21]. Based on the SC principle, the symbol vector transmitted is given by $\mathbf{x}^{(i)} = \mathbf{x}_0^{(i)} + \mathbf{x}_1^{(i)} + ... + \mathbf{x}_K^{(i)}$, where $\mathbf{x}_0^{(i)}$ for all $i$ encodes a multicast message intended to all users and $\mathbf{x}_k^{(i)}$ for all $i$ encodes a unicast message intended to user $k$ only.

Users may be arranged in the decreasing order of their channel SNRs to determine the order of the SIC process. However, with OFDM the user ordering may be different in each subchannel. Due to this, we let $\pi^{(i)}$ be the permutation of the users such that $\pi^{(i)}(k)$ is the index of user that has $k^{\text{th}}$ largest channel SNR in subchannel $i$ (i.e., $\sigma_{i,\pi^{(i)}(1)}^2 > \sigma_{i,\pi^{(i)}(2)}^2 > \cdots > \sigma_{i,\pi^{(i)}(K)}^2$ for all $i$)[3]. We shall simply refer to the user $\pi^{(i)}(1)$ as the *strongest user* in subchannel $i$.

In the degraded broadcast channel, signal received by user $\pi^{(i)}(k+1)$ is a (stochastically) degraded version of the signal received by user $\pi^{(i)}(k)$ for all $k = 1, \ldots, K-1$. Thus, if user $\pi^{(i)}(k+1)$ can decode its $(k+1)^{\text{th}}$ unicast signal, then user $\pi^{(i)}(k)$ can also decode (and cancel) it. This observation

---

[3] The probability of the channel SNRs of any two users being exactly equal is zero when channels are independently drawn from a continuous distribution. In practice, one can break any ties by subtracting a small $\varepsilon > 0$ from one of the channel SNRs with negligible penalty in the achieved rates.

motivates the SIC decoding process as follows.

Upon receiving the signal (1), user $k$ first decodes the multicast signal while treating all unicast signals as noise. Once the user cancels the decoded multicast signal from (1), it then proceeds to decode and cancel unicast signals sequentially starting from the signal of the weakest user $\pi^{(i)}(K)$, then the signal of the next weakest user $\pi^{(i)}(K-1)$, and so on, until the desired user signal is decoded. In each step, all signals that have not been decoded until that stage are treated as noise.

### B. Achievable rate region

A Gaussian signaling over a MIMO channel $\mathbf{H}$ with a (Gaussian) interfering signal has the ergodic capacity [41]:

$$\mathbb{E}_{\mathbf{H}}\left[\log\det\left(\mathbf{I}+\left(\sigma^2\mathbf{I}+\mathbf{H}\mathbf{Q}_j\mathbf{H}^\dagger\right)^{-1}\mathbf{H}\mathbf{Q}_i\mathbf{H}^\dagger\right)\right],$$

where $\mathbf{Q}_i$ and $\mathbf{Q}_j$ are the covariance matrices of the signal and the interferer, respectively, and $\sigma^2$ is the variance of AWGN. The term $\left(\sigma^2\mathbf{I}+\mathbf{H}\mathbf{Q}_j\mathbf{H}^\dagger\right)$ is the overall covariance matrix of the noise plus the interference at the receiver. The expected value is carried over the probability distribution of $\mathbf{H}$, and due to the statistical CSIT assumption, $\mathbf{Q}_i$ cannot be a function of the individual realizations of $\mathbf{H}$.

We denote $R_0$ as the rate of the multicast message and $R_k$ for $k \in \mathcal{K}$ as the rate of the unicast message of user $k$. In decoding the multicast signal, all unicast signals act as interference, while in decoding the unicast signal, only a subset of the unicast signals that cannot be decoded in the SIC process act as interference. Furthermore, $R_0$ needs to be set to the minimum of the individual user rates to ensure that all users can decode the multicast signal. Incorporating these constraints in each of the $M$ parallel OFDM channels, the overall achievable rate region may be expressed as the union of the following rate tuple ($R_0$, $R_1$, …, $R_K$) in bps/Hz:

$$R_0 \leq \min_{k \in \mathcal{K}}\left\{\sum_{i=1}^M \eta_i \mathbb{E}\left[\log\det\left(\mathbf{I}+\left(\eta_i\mathbf{I}+\sum_{j=1}^K\mathbf{H}_k^{(i)}\mathbf{Q}_j^{(i)}\mathbf{H}_k^{(i)\dagger}\right)^{-1}\right.\right.\right.$$
$$\left.\left.\left.\cdot\mathbf{H}_k^{(i)}\mathbf{Q}_0^{(i)}\mathbf{H}_k^{(i)\dagger}\right)\right]\right\},$$

$$R_k \leq \sum_{i=1}^M \eta_i \mathbb{E}\left[\log\det\left(\mathbf{I}+\left(\eta_i\mathbf{I}+\sum_{j\in\mathcal{G}_k^{(i)}}\mathbf{H}_k^{(i)}\mathbf{Q}_j^{(i)}\mathbf{H}_k^{(i)\dagger}\right)^{-1}\right.\right.$$
$$\left.\left.\cdot\mathbf{H}_k^{(i)}\mathbf{Q}_k^{(i)}\mathbf{H}_k^{(i)\dagger}\right)\right], \quad k \in \mathcal{K}, \tag{3}$$

where $\mathbf{Q}_0^{(i)}$ is the covariance matrix of $\mathbf{x}_0^{(i)}$ (multicast), $\mathbf{Q}_k^{(i)}$ is the covariance matrix of $\mathbf{x}_k^{(i)}$ (unicast), and $\mathcal{G}_k^{(i)}$ denotes the set of users whose channel SNRs in subchannel $i$ are larger than that of user $k$, i.e., $\mathcal{G}_k^{(i)} = \left\{\forall j \in \mathcal{K}: \sigma_{i,j}^2 > \sigma_{i,k}^2\right\}$, hence the unicast signals intended for those users remain as interference at user $k$.

### III. Ergodic Weighted Sum-Rate Maximization Problem

In this work, we consider an ergodic weighted sum-rate (WSR) maximization problem $\max\left\{\mu R_0 + \sum_{k\in\mathcal{K}} R_k\right\}$ with non-negative rate reward $\mu$ given towards the multicast. Here, $\mu$ controls how much priority is given towards the multicast and unicast messages: Increasing $\mu$ beyond 1 gives higher priority to the multicast message and leads to a boundary point on the achievable rate region with higher $R_0$. It is worth noting that the problem formulation with $\mu$ results in more general framework than the (unweighted) sum-rate formulation, and includes, as a special case, the maximum network throughput when $\mu$ is equal to the number of users, $K$.

We impose an average power constraint and a minimum multicast rate constraint $R_0 \geq R_{0,\min}$, but the latter can be satisfied (if feasible) by incrementing the rate reward $\mu$ until the multicast rate constraint is satisfied with an equality – hence the minimum multicast rate constraint may be dropped without loss of generality while keeping $\mu$ arbitrary.

### A. Reformulated problem and structure of the optimal solution

Let the entries of $\mathbf{H} \in \mathbb{C}^{n_R \times n_T}$ be distributed as $\mathcal{CN}(0,1)$. Using the fact that $\mathbf{H}_k^{(i)}$ and $\sigma_{\mathbf{H}_k^{(i)}}\mathbf{H}$ have the same distribution and using the definition $\sigma_{i,k}^2 = \sigma_{\mathbf{H}_k^{(i)}}^2/\eta_i$, we define

$$R_{0,k}^{(i)} \triangleq \mathbb{E}\left[\log\det\left(\mathbf{I}+\left(\mathbf{I}+\sigma_{i,k}^2\sum_{j=1}^K\mathbf{H}\mathbf{Q}_j^{(i)}\mathbf{H}^\dagger\right)^{-1}\sigma_{i,k}^2\mathbf{H}\mathbf{Q}_0^{(i)}\mathbf{H}^\dagger\right)\right], \tag{4}$$

which represents the rate of the multicast message in subchannel $i$ so that it may be decoded by user $k$. Similarly, the rate of the unicast message of $k^{\text{th}}$ strongest user, $\pi^{(i)}(k)$, in subchannel $i$ is defined as

$$R_{\pi^{(i)}(k)}^{(i)} \triangleq \mathbb{E}\left[\log\det\left(\mathbf{I}+\left(\mathbf{I}+\sigma_{i,\pi^{(i)}(k)}^2\sum_{j=1}^{k-1}\mathbf{H}\mathbf{Q}_{\pi^{(i)}(j)}^{(i)}\mathbf{H}^\dagger\right)^{-1}\right.\right.$$
$$\left.\left.\cdot\sigma_{i,\pi^{(i)}(k)}^2\mathbf{H}\mathbf{Q}_{\pi^{(i)}(k)}^{(i)}\mathbf{H}^\dagger\right)\right]. \tag{5}$$

Then, the WSR maximization problem is formulated as:

$$\max_{\left\{\mathbf{Q}_0^{(i)},\mathbf{Q}_k^{(i)}\right\}\succeq 0, \forall i, \forall k}\left\{\mu\min_{k\in\mathcal{K}}\left\{\sum_{i=1}^M \eta_i R_{0,k}^{(i)}\right\} + \sum_{k=1}^K\left(\sum_{i=1}^M \eta_i R_{\pi^{(i)}(k)}^{(i)}\right)\right\}, \tag{6}$$

subject to the average power constraint

$$\sum_{i\in\mathcal{M}} tr\left(\mathbf{Q}_0^{(i)} + \mathbf{Q}_1^{(i)} + \cdots + \mathbf{Q}_K^{(i)}\right) \leq P_t, \tag{7}$$

where $P_t$ is the total available power.

The WSR problem is generally nonconvex in the input covariances $\mathbf{Q}$ and its solutions are not available in a closed form, unless $\mu$ is less than 1 (for which the optimal $\mathbf{Q}_0^{(i)}$ for multicast are all zeros) or as $\mu$ tends to infinity (for which the optimal $\mathbf{Q}_k^{(i)}$ for unicast are all zeros). For $\mu > 1$, we identify a range of $\mu$ for which the optimal $\mathbf{Q}$'s are nonzero − hence the joint transmission of multicast and unicast is optimal − and structures of those $\mathbf{Q}$'s.

In what follows, we first reformulate the WSR problem into a more convenient form for algorithmic purposes. We note that the WSR problem is a *max-min* problem of the following form:

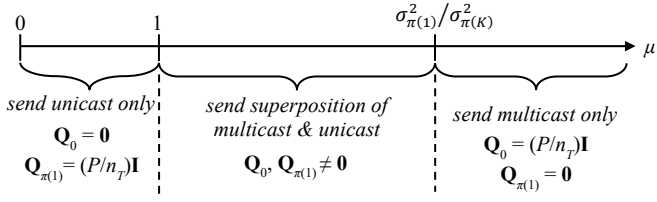

Fig. 4. Regions of $\mu$ (multicast rate reward) and the respective optimal transmission policies in the case of a single subchannel ($M = 1$). $\sigma^2_{\pi(1)}$ and $\sigma^2_{\pi(K)}$ denote the best user and the worst user channel SNRs, respectively.

$$\max_{\boldsymbol{p} \in \mathcal{P}} \left\{ \min \left\{ G_1(\boldsymbol{p}), G_2(\boldsymbol{p}), \cdots, G_n(\boldsymbol{p}) \right\} + G_{n+1}(\boldsymbol{p}) \right\}, \quad (8)$$

where $G_1(\boldsymbol{p}), \ldots, G_{n+1}(\boldsymbol{p})$ are real continuous functions of $\boldsymbol{p}$. This class of max-min problem may be solved by the method introduced in [42], summarized in the following lemma.

*Lemma 1* [Proposition 1, [42]]: Let $\boldsymbol{\alpha} \triangleq [\alpha_1, \alpha_2, \cdots, \alpha_n]$ with non-negative $\alpha_i$'s and $\sum_{i=1}^{n} \alpha_i = 1$. Furthermore, let

$$G(\boldsymbol{\alpha}, \boldsymbol{p}) \triangleq \sum_{i=1}^{n} \alpha_i G_i(\boldsymbol{p}) + G_{n+1}(\boldsymbol{p}) = \sum_{i=1}^{n} \alpha_i \left( G_i(\boldsymbol{p}) + G_{n+1}(\boldsymbol{p}) \right)$$

and $V(\boldsymbol{\alpha}) \triangleq \max_{\boldsymbol{p} \in \mathcal{P}} G(\boldsymbol{\alpha}, \boldsymbol{p}) = G(\boldsymbol{\alpha}, \boldsymbol{p}(\boldsymbol{\alpha}))$,

where $\boldsymbol{p}(\boldsymbol{\alpha})$ denotes the $\boldsymbol{p}$ that maximizes $G(\boldsymbol{\alpha}, \boldsymbol{p})$ for a fixed $\boldsymbol{\alpha}$. Suppose $\boldsymbol{\alpha}^*$ is a minimizer of $V$, i.e., $V(\boldsymbol{\alpha}^*) = \min_{\boldsymbol{\alpha}} V(\boldsymbol{\alpha})$. Then, $\boldsymbol{p}(\boldsymbol{\alpha}^*)$ is the solution to the *max-min* problem in (8).

This lemma reformulates the max-min problem as a min-max problem, where the minimization can be efficiently carried out using convex optimization methods due to $V(\boldsymbol{\alpha})$ being continuous and convex in $\boldsymbol{\alpha}$. Using Lemma 1, our WSR optimization problem can be reformulated as the following:

$$\min_{\boldsymbol{\mu}} \left\{ \max_{\{\mathbf{Q}_0^{(i)}, \mathbf{Q}_k^{(i)}\} \succeq \mathbf{0}, \forall i, \forall k} \left\{ \sum_{i=1}^{M} \eta_i \sum_{k=1}^{K} \left( \mu_k R_{0,k}^{(i)} + R_{\pi^{(i)}(k)}^{(i)} \right) \right\} \right\} \quad (9)$$

s.t. (7)

for a non-negative weight vector $\boldsymbol{\mu} \triangleq [\mu_1, \ldots, \mu_K]$ that satisfies $\sum_{k \in \mathcal{K}} \mu_k = \mu$. The following shows the structure of the optimal input covariance matrices $\mathbf{Q}$ that solves (9), which depends on the multicast rate reward $\mu$.

*Theorem 1*: Let $P^{(i)} \triangleq tr\left(\mathbf{Q}_0^{(i)} + \sum_{k \in \mathcal{K}} \mathbf{Q}_k^{(i)}\right)$ denote the power allocated to subchannel $i$. The optimal covariance matrices that solve the ergodic WSR maximization problem satisfy the following four conditions:

1. $\mathbf{Q}_{\pi^{(i)}(2)}^{(i)} = \mathbf{Q}_{\pi^{(i)}(3)}^{(i)} = \ldots = \mathbf{Q}_{\pi^{(i)}(K)}^{(i)} = \mathbf{0} \quad \forall i$.
2. $\mathbf{Q}_0^{(i)}, \mathbf{Q}_{\pi^{(i)}(1)}^{(i)} \quad \forall i$ are sufficiently non-negative diagonal.
3. When $\mu \leq 1$, $\mathbf{Q}_0^{(i)} = \mathbf{0}$ and $\mathbf{Q}_{\pi^{(i)}(1)}^{(i)} = \left(P^{(i)}/n_T\right)\mathbf{I} \quad \forall i$.
4. When $\mu > 1$,
   4.1. if $\sum_{k \in \mathcal{K}} \mu_k \sigma_{i,k}^2 \geq \sigma_{i,\pi^{(i)}(1)}^2$, then $\mathbf{Q}_0^{(i)} = \left(P^{(i)}/n_T\right)\mathbf{I}$ and $\mathbf{Q}_{\pi^{(i)}(1)}^{(i)} = \mathbf{0} \quad \forall i$.
   4.2. if $\sum_{k \in \mathcal{K}} \mu_k \sigma_{i,k}^2 < \sigma_{i,\pi^{(i)}(1)}^2$, then $\mathbf{Q}_{\pi^{(i)}(1)}^{(i)} \neq \mathbf{0} \quad \forall i$.

*Proof*: A proof is provided in subsection III.B. ∎

The condition 1 implies that it is generally optimal to send in each subchannel both the multicast message and a *single* unicast message of the strongest user only. In other words, *greedy user selection* based on largest channel SNR and *two-layered superposition* of multicast and single unicast are WSR optimal (similar observations were made in the SISO OFDMA systems [35]). The condition 2 implies that no correlations are needed in the space (or across the antennas). The condition 3 indicates that it is optimal to send only the unicasts in all subchannels when $\mu \leq 1$, whereas in the case of $\mu > 1$, the condition 4 shows when to send the multicast only and when to send the superposition of multicast and unicast.

The conditions 3 and 4 become more explicit in the case of a single subchannel ($M = 1$). With only one channel, there exists a clear sense of who the worst user is, say $\pi(K)^{\text{th}}$ user, to whom the multicast rate will target. Consequently, the optimal weight vector $\boldsymbol{\mu}$ becomes a vector with a single nonzero element with $\mu_{\pi(K)} = \mu$, and the condition $\sum_{k \in \mathcal{K}} \mu_k \sigma_{i,k}^2 < \sigma_{i,\pi^{(i)}(1)}^2$ reduces to simply $\mu < \sigma_{\pi(1)}^2 / \sigma_{\pi(K)}^2$, i.e., ratio of the best and the worst user channel SNRs. Fig. 4 summarizes the regions of $\mu$ and the respective optimal transmission policies, showing that both multicast and unicast should be sent when the multicast rate reward outweighs the unicast rate reward but still less than the ratio of the best and the worst user channel SNRs.

Theorem 1 narrows down the only unknown case to when $\sum_{k \in \mathcal{K}} \mu_k \sigma_{i,k}^2 < \sigma_{i,\pi^{(i)}(1)}^2$, in which the closed form solutions to $\mathbf{Q}_0^{(i)}$ and $\mathbf{Q}_{\pi^{(i)}(1)}^{(i)}$ are unknown. In the case of isotropic Rayleigh fading, all spatial directions are equally good in average and hence equally distributing the power across the antenna array seems intuitively optimal. However, how the multicast and the unicast should optimally share the spatial directions is an open problem. We show next that the equal power allocations across the antenna arrays for both $\mathbf{Q}_0^{(i)}$ and $\mathbf{Q}_{\pi^{(i)}(1)}^{(i)} \quad \forall i$ are at least guaranteed to be *locally optimal*.

*Theorem 2*: Let $P^{(i)} = tr\left(\mathbf{Q}_0^{(i)} + \mathbf{Q}_{\pi^{(i)}(1)}^{(i)}\right)$, $P_1^{(i)} \triangleq tr\left(\mathbf{Q}_{\pi^{(i)}(1)}^{(i)}\right)$, both non-negative. Then for all $i$, $\mathbf{Q}_0^{(i)} = \frac{1}{n_T}\left(P^{(i)} - P_1^{(i)}\right) \cdot \mathbf{I}$ and $\mathbf{Q}_{\pi^{(i)}(1)}^{(i)} = \frac{1}{n_T} P_1^{(i)} \cdot \mathbf{I}$ are local optimal solutions of the ergodic WSR maximization problem.
*Proof*: See Appendix A. ∎

It should be noted that while the structure of the optimal input greatly simplifies the design of the joint multicast and unicast transmissions, such design alone does not provide a QoS guarantee for the unicast services. If a certain user has mediocre channels across all $M$ subchannels, no unicast message will be delivered to the user (while in contrast, the multicast message is always delivered to all users). Attaining fairness or meeting certain QoS requirements between unicast services can be achieved via employing an additional scheduling algorithm such as round-robin or proportional fair, but these will necessarily deviate the performance away from the maximum WSR operating point. Allowing different rate rewards among the unicast traffics can also attain fairness, but the optimal solution generally involves more than two layers of SC, which greatly complicates the optimal resource allocation and increases the effect of error propagation in SIC in practical systems with moderate to large number of users.

### B. Proof of Theorem 1

We define
$$\Phi_k^{(i)}(\mathbf{Q}) \triangleq \mathbb{E}_\mathbf{H}\left[\log\det\left(\mathbf{I} + \sigma_{i,k}^2 \mathbf{H}\mathbf{Q}\mathbf{H}^\dagger\right)\right]. \quad (10)$$

The following log-determinant identity will be useful:
$$\begin{aligned}
&\log\det\left(\mathbf{I} + \left(\mathbf{I} + \mathbf{H}\mathbf{A}\mathbf{H}^\dagger\right)^{-1}\mathbf{H}\mathbf{B}\mathbf{H}^\dagger\right) \\
&= \log\det\left(\mathbf{I} + \mathbf{H}(\mathbf{A}+\mathbf{B})\mathbf{H}^\dagger\right) - \log\det\left(\mathbf{I} + \mathbf{H}\mathbf{A}\mathbf{H}^\dagger\right).
\end{aligned} \quad (11)$$

In this proof, we abbreviate $\mathbf{Q}_{\pi^{(i)}(k)}^{(i)}$ by $\mathbf{Q}_k^{(i)}$ for all $k$ unless otherwise noted. We sequentially prove the 4 conditions of the theorem below.

*Proof of the condition 1:* We define the objective function (9) in the $i^{\text{th}}$ channel as
$$\begin{aligned}
&F\left(\mathbf{Q}_0^{(i)}, \mathbf{Q}_1^{(i)}, \cdots, \mathbf{Q}_K^{(i)}\right) \\
&\triangleq \sum_{k=1}^K \mu_k R_{0,k}^{(i)}\left(\mathbf{Q}_0^{(i)}, \mathbf{Q}_1^{(i)}, \cdots, \mathbf{Q}_K^{(i)}\right) + \sum_{k=1}^K R_{\pi^{(i)}(k)}^{(i)}\left(\mathbf{Q}_1^{(i)}, \cdots, \mathbf{Q}_k^{(i)}\right),
\end{aligned} \quad (12)$$

where $R_{0,k}^{(i)}$ and $R_{\pi^{(i)}(k)}^{(i)}$ are as in (4) and (5), respectively, which are expressed as functions of $\mathbf{Q}$'s to signify their dependence. We wish to show that $F$ can only increase by decreasing the trace of $\mathbf{Q}_2^{(i)}$. To this end, for a fixed set of $\mathbf{Q}$'s, let
$$\mathbf{Q}_2^{(i)}(\Delta) \triangleq \mathbf{Q}_2^{(i)} - \Delta\frac{\mathbf{Q}_2^{(i)}}{tr(\mathbf{Q}_2^{(i)})} \text{ and } \mathbf{Q}_1^{(i)}(\Delta) \triangleq \mathbf{Q}_1^{(i)} + \Delta\frac{\mathbf{Q}_2^{(i)}}{tr(\mathbf{Q}_2^{(i)})}$$

for $0 \leq \Delta \leq tr(\mathbf{Q}_2^{(i)})$. Note that $\mathbf{Q}_1^{(i)}(\Delta) + \mathbf{Q}_2^{(i)}(\Delta) = \mathbf{Q}_1^{(i)} + \mathbf{Q}_2^{(i)}$. Noting that $R_{0,k}^{(i)}$ depends only on $\mathbf{Q}_0^{(i)}$ and $\mathbf{Q}_1^{(i)} + \cdots + \mathbf{Q}_K^{(i)}$ (but not on the individual $\mathbf{Q}_k^{(i)}$'s), $R_{0,k}^{(i)}\left(\mathbf{Q}_1^{(i)}(\Delta), \mathbf{Q}_2^{(i)}(\Delta)\right)$ is independent of $\Delta$ for all $k$. Similarly, since $R_{\pi^{(i)}(k)}^{(i)}$ depends only on $\mathbf{Q}_1^{(i)} + \cdots + \mathbf{Q}_{k-1}^{(i)}$ and $\mathbf{Q}_k^{(i)}$, $R_{\pi^{(i)}(k)}^{(i)}\left(\mathbf{Q}_1^{(i)}(\Delta), \mathbf{Q}_2^{(i)}(\Delta)\right)$ is independent of $\Delta$ for all $k \geq 3$. Therefore,
$$\begin{aligned}
&\frac{\partial}{\partial\Delta}F\left(\mathbf{Q}_1^{(i)}(\Delta), \mathbf{Q}_2^{(i)}(\Delta)\right) \\
&= \frac{\partial}{\partial\Delta}\left(R_{\pi^{(i)}(1)}^{(i)}\left(\mathbf{Q}_1^{(i)}(\Delta)\right) + R_{\pi^{(i)}(2)}^{(i)}\left(\mathbf{Q}_1^{(i)}(\Delta), \mathbf{Q}_2^{(i)}(\Delta)\right)\right).
\end{aligned}$$

The last term, $R_{\pi^{(i)}(2)}^{(i)}$, can be further simplified as, using (11),
$$\begin{aligned}
&R_{\pi^{(i)}(2)}^{(i)}\left(\mathbf{Q}_1^{(i)}(\Delta), \mathbf{Q}_2^{(i)}(\Delta)\right) \\
&= \Phi_{\pi^{(i)}(2)}^{(i)}\left(\mathbf{Q}_1^{(i)}(\Delta) + \mathbf{Q}_2^{(i)}(\Delta)\right) - \Phi_{\pi^{(i)}(2)}^{(i)}\left(\mathbf{Q}_1^{(i)}(\Delta)\right),
\end{aligned}$$

where the first term in the RHS is independent of $\Delta$. Therefore,
$$\begin{aligned}
\frac{\partial F}{\partial\Delta} &= \frac{\partial}{\partial\Delta}\left(\Phi_{\pi^{(i)}(1)}^{(i)}\left(\mathbf{Q}_1^{(i)}(\Delta)\right) - \Phi_{\pi^{(i)}(2)}^{(i)}\left(\mathbf{Q}_1^{(i)}(\Delta)\right)\right) \\
&\stackrel{(a)}{=} \mathbb{E}\left[tr\left(\left(\sigma_{i,\pi^{(i)}(1)}^{-2}\mathbf{I} + \mathbf{H}\mathbf{Q}_1^{(i)}(\Delta)\mathbf{H}^\dagger\right)^{-1}\right.\right. \\
&\qquad \left.\left. - \left(\sigma_{i,\pi^{(i)}(2)}^{-2}\mathbf{I} + \mathbf{H}\mathbf{Q}_1^{(i)}(\Delta)\mathbf{H}^\dagger\right)^{-1}\right)\frac{1}{tr(\mathbf{Q}_2^{(i)})}\mathbf{H}\mathbf{Q}_2^{(i)}\mathbf{H}^\dagger\right] \\
&\stackrel{(b)}{\geq} 0,
\end{aligned}$$

where (a) is obtained by matrix calculus (e.g., eq. (46) in [43]) and (b) is due to $\sigma_{\pi^{(i)}(1)}^2 \geq \sigma_{\pi^{(i)}(2)}^2$ and the trace of product of two positive semidefinite matrices is non-negative. This implies that the objective function can only increase by increasing $\Delta$, leading to $\mathbf{Q}_2^{(i)} = \mathbf{0}$ being the optimal solution.

Setting $\mathbf{Q}_2^{(i)} = \mathbf{0}$ and repeating the same procedure for $\mathbf{Q}_3^{(i)}$ leads to $\mathbf{Q}_3^{(i)} = \mathbf{0}$ being optimal. The same procedure can continue until $\mathbf{Q}_K^{(i)}$, leading to $\mathbf{Q}_2^{(i)} = \cdots = \mathbf{Q}_K^{(i)} = \mathbf{0}$, $\forall i$ being optimal. This completes the proof of the condition 1. ∎

*Proof of the condition 2:* Setting $\mathbf{Q}_2^{(i)} = \cdots = \mathbf{Q}_K^{(i)} = \mathbf{0}$ and applying the log-determinant identity (11), the maximization problem of (9) reduces to
$$\begin{aligned}
\max_{\{\mathbf{Q}_0^{(i)}, \mathbf{Q}_1^{(i)}\}\succeq \mathbf{0}, \forall i} &\left\{\sum_{i=1}^M \eta_i\left(\sum_{k=1}^K \mu_k \Phi_k^{(i)}\left(\mathbf{Q}_0^{(i)} + \mathbf{Q}_1^{(i)}\right)\right.\right. \\
&\left.\left. - \sum_{k=1}^K \mu_k \Phi_k^{(i)}\left(\mathbf{Q}_1^{(i)}\right) + \Phi_{\pi^{(i)}(1)}^{(i)}\left(\mathbf{Q}_1^{(i)}\right)\right)\right\} \\
\text{s.t.} \quad &\sum_{i\in\mathcal{M}} tr\left(\mathbf{Q}_0^{(i)} + \mathbf{Q}_1^{(i)}\right) \leq P_t.
\end{aligned} \quad (13)$$

Let $\mathbf{Q}_0^{(i)}$ and $\mathbf{Q}_1^{(i)}$ be the optimal solutions. Because $\mathbf{H}$ is isotropic and statistically invariant to unitary transformations, $\Phi_k^{(i)}(\mathbf{Q}) = \Phi_k^{(i)}(\mathbf{U}\mathbf{Q}\mathbf{U}^\dagger)$ for any unitary $\mathbf{U}$. Thus, the objective function is unchanged if we replace $\mathbf{Q}_0^{(i)} + \mathbf{Q}_1^{(i)}$ by the diagonal matrix $\mathbf{D}^{(i)}$ whose diagonal entries correspond to its eigenvalues, and $\mathbf{Q}_1^{(i)}$ by the diagonal $\mathbf{\Lambda}^{(i)}$ with its entries corresponding to its eigenvalues. That is, $\Phi_k^{(i)}\left(\mathbf{Q}_0^{(i)}+\mathbf{Q}_1^{(i)}\right)=\Phi_k^{(i)}\left(\mathbf{D}^{(i)}\right)$ and $\Phi_k^{(i)}\left(\mathbf{Q}_1^{(i)}\right)=\Phi_k^{(i)}\left(\mathbf{\Lambda}^{(i)}\right)$. Furthermore, since $\Phi_k^{(i)}$ is invariant to any unitary transformations, we may choose the same set of eigenvectors for both $\mathbf{Q}_0^{(i)}+\mathbf{Q}_1^{(i)}$ and $\mathbf{Q}_1^{(i)}$ without changing the objective function. Hence, it suffices to choose $\mathbf{Q}_0^{(i)}+\mathbf{Q}_1^{(i)}=\mathbf{U}^{(i)}\mathbf{D}^{(i)}\mathbf{U}^{(i)\dagger}$ and $\mathbf{Q}_1^{(i)}=\mathbf{U}^{(i)}\mathbf{\Lambda}^{(i)}\mathbf{U}^{(i)\dagger}$ with the same set of unitary matrices $\mathbf{U}^{(i)}$, and we conclude that both $\mathbf{Q}_0^{(i)}$ and $\mathbf{Q}_1^{(i)}$ are sufficient to be non-negative diagonal. This proves the condition 2. ∎

*Proof of the condition 3:* When $\mu \leq 1$, the procedure shown in the proof of the condition 1 can be applied once more on $\mathbf{Q}_0^{(i)}$, which leads to $\mathbf{Q}_0^{(i)}=\mathbf{0}$ being optimal, leaving only $\mathbf{Q}_1^{(i)}$ nonzero. It is well known that in the case of single unicast message, the optimal $\mathbf{Q}_1^{(i)}$ is a scalar matrix [41], i.e., $\mathbf{Q}_1^{(i)}=\frac{P^{(i)}}{n_T}\mathbf{I}$. This completes the proof of the condition 3. ∎

*Proof of the condition 4.1:* Assume $\mu > 1$ and $\sum_{k\in\mathcal{K}}\mu_k\sigma_{i,k}^2 \geq \sigma_{i,\pi^{(i)}(1)}^2$. Our goal is to show that the objective function of (13) is at most $\sum_{i\in\mathcal{M}}\eta_i\sum_{k\in\mathcal{K}}\mu_k\Phi_k^{(i)}\left(\frac{P^{(i)}}{n_T}\mathbf{I}\right)$, which is achieved by $\mathbf{Q}_0^{(i)}=\frac{P^{(i)}}{n_T}\mathbf{I}$ and $\mathbf{Q}_1^{(i)}=\mathbf{0}$. First note that the first term of the objective function of (13) is upper-bounded by choosing $\mathbf{Q}_0^{(i)}+\mathbf{Q}_1^{(i)}=\frac{P^{(i)}}{n_T}\mathbf{I}$ (see [41]), i.e.,

$$\sum_{k=1}^K \mu_k\Phi_k^{(i)}\left(\mathbf{Q}_0^{(i)}+\mathbf{Q}_1^{(i)}\right) \leq \sum_{k=1}^K \mu_k\Phi_k^{(i)}\left(\frac{P^{(i)}}{n_T}\mathbf{I}\right).$$

As for the remaining terms, we define

$$G\left(\mathbf{Q}_1^{(i)}\right) \triangleq -\sum_{k=1}^K \mu_k\Phi_k^{(i)}\left(\mathbf{Q}_1^{(i)}\right) + \Phi_{\pi^{(i)}(1)}^{(i)}\left(\mathbf{Q}_1^{(i)}\right).$$

Let $\mathbf{Q}_1^{(i)}(\Delta) \triangleq \Delta \cdot \mathbf{Q}_1^{(i)}/tr\left(\mathbf{Q}_1^{(i)}\right)$ for $\Delta \geq 0$. Then,

$$\frac{\partial G\left(\mathbf{Q}_1^{(i)}(\Delta)\right)}{\partial \Delta} =$$
$$-\sum_{k=1}^K \mu_k\sigma_{i,k}^2 \mathbb{E}\left[tr\left(\left(\mathbf{I}+\sigma_{i,k}^2\mathbf{HQ}_1^{(i)}(\Delta)\mathbf{H}^\dagger\right)^{-1}\frac{1}{tr(\mathbf{Q}_1^{(i)})}\mathbf{HQ}_1^{(i)}\mathbf{H}^\dagger\right)\right]$$
$$+\sigma_{i,\pi^{(i)}(1)}^2 \mathbb{E}\left[tr\left(\left(\mathbf{I}+\sigma_{i,\pi^{(i)}(1)}^2\mathbf{HQ}_1^{(i)}(\Delta)\mathbf{H}^\dagger\right)^{-1}\frac{1}{tr(\mathbf{Q}_1^{(i)})}\mathbf{HQ}_1^{(i)}\mathbf{H}^\dagger\right)\right] \quad (14)$$
$$\overset{(a)}{\leq} \sum_{k=1}^K \mu_k\sigma_{i,k}^2 \left\{ -\mathbb{E}\left[tr\left(\left(\mathbf{I}+\sigma_{i,k}^2\mathbf{HQ}_1^{(i)}(\Delta)\mathbf{H}^\dagger\right)^{-1}\frac{1}{tr(\mathbf{Q}_1^{(i)})}\mathbf{HQ}_1^{(i)}\mathbf{H}^\dagger\right)\right] \right.$$
$$\left. +\mathbb{E}\left[tr\left(\left(\mathbf{I}+\sigma_{i,\pi^{(i)}(1)}^2\mathbf{HQ}_1^{(i)}(\Delta)\mathbf{H}^\dagger\right)^{-1}\frac{1}{tr(\mathbf{Q}_1^{(i)})}\mathbf{HQ}_1^{(i)}\mathbf{H}^\dagger\right)\right] \right\}$$
$$\overset{(b)}{\leq} 0,$$

where (a) is due to $\sum_{k\in\mathcal{K}}\mu_k\sigma_{i,k}^2 \geq \sigma_{i,\pi^{(i)}(1)}^2$ and (b) is due to $\sigma_{i,\pi^{(i)}(1)}^2 > \sigma_{i,k}^2$ for all $k\neq \pi^{(i)}(1)$. This implies that $G$ is decreasing in $\Delta$, and combined with the fact that $G=0$ at $\Delta=0$, it is optimal to choose $\mathbf{Q}_1^{(i)}=\mathbf{0}$ and $\mathbf{Q}_0^{(i)}=\frac{P^{(i)}}{n_T}\mathbf{I}$ to maximize the objective function. This proves the condition 4.1. ∎

*Proof of the condition 4.2:* Assume $\mu > 1$ and $\sum_{k\in\mathcal{K}}\mu_k\sigma_{i,k}^2 < \sigma_{i,\pi^{(i)}(1)}^2$. Evaluating (14) at $\Delta = 0$ yields

$$\left.\frac{\partial G}{\partial \Delta}\right|_{\Delta=0} = \left(-\sum_{k=1}^K \mu_k\sigma_{i,k}^2 + \sigma_{i,\pi^{(i)}(1)}^2\right)\mathbb{E}\left[tr\left(\frac{1}{tr(\mathbf{Q}_1^{(i)})}\mathbf{HQ}_1^{(i)}\mathbf{H}^\dagger\right)\right],$$

which is strictly positive, meaning $G$ is initially increasing at $\Delta = 0$. Hence, the condition $\sum_{k\in\mathcal{K}}\mu_k\sigma_{i,k}^2 = \sigma_{i,\pi^{(i)}(1)}^2$ is a sharp boundary between the regions when only multicast should be sent ($\Delta = 0$) and when unicast should be sent along with multicast ($\Delta > 0$). While unneeded for this proof, it is interesting to see that evaluating (14) as $\Delta \to \infty$ yields

$$\left.\frac{\partial G}{\partial \Delta}\right|_{\Delta\to\infty} = \Delta^{-1}n_R\left(-\mu+1\right) < 0,$$

which implies that $G$ attains a maximum at some finite $\Delta > 0$. This completes the proof of the condition 4.2. ∎

### C. Degrees-of-freedom analysis

Since it is optimal to send one unicast message per subchannel under the statistical CSIT assumption (cf., Theorem 1), the DoF of the unicast message linearly scales with the number of subchannels but not with the number of users. As for the multicast, the DoF linearly scales with both $M$ and $K$ since all subchannels are used and all users receive the message.

Formal analysis of the DoF is shown below. The DoF (or multiplexing gain) of user $k$ in bits/second/Hz is defined as

$$d_k \triangleq \lim_{P_t\to\infty}\frac{R_0(P_t)+R_k(P_t)}{\log P_t}, \quad (15)$$

and the sum DoF is

$$d_s \triangleq \lim_{P_t\to\infty}\sum_{k\in\mathcal{K}}d_k = \lim_{P_t\to\infty}\frac{\sum_{k\in\mathcal{K}}(R_0(P_t)+R_k(P_t))}{\log P_t}, \quad (16)$$

where $R_0(P_t)$ is the multicast rate and $R_k(P_t)$ is the unicast rate of user $k$, expressed as functions of total available power $P_t$. The sum rate in the asymptotic SNR region is first characterized below.

*Proposition 1:* Let $n \triangleq \min(n_T, n_R)$. The sum rate $R_s(P_t) \triangleq \sum_{k\in\mathcal{K}}(R_0(P_t)+R_k(P_t))$ in the limit $P_t \to \infty$ may be expressed as

$$\lim_{P_t \to \infty} R_s(P_t) = K\left(n \log P_t - n' \sum_{i \in \mathcal{I}} \eta_i \log P^{(i)}_{\pi^{(i)}(1)}\right) \\ + n' \sum_{i \in \mathcal{I}} \eta_i \log P^{(i)}_{\pi^{(i)}(1)} + o(1), \quad (17)$$

where $n' \leq n$ is the number of spatial dimensions used by the unicast, $\mathcal{I}$ is the set of OFDM subchannels where unicast is sent with power that scales with $P_t$, e.g., $P^{(i)}_{\pi^{(i)}(1)} = \alpha_i P_t$ with $0 < \alpha_i \leq 1$, and $o(1)$ includes all terms that do not scale with $P_t$.

*Proof*: See Appendix B. ∎

Consequently, we can derive the sum DoF in three scenarios of interest:

1. Giving all power to multicast (i.e., $n' = 0$, $P^{(i)}_{\pi^{(i)}(1)} = 0$, $\mathcal{I} = \{\}$) yields $d_s = Kn$. This is the full DoF.

2. Giving all power to the unicast (i.e., $n' = n$, $P^{(i)}_{\pi^{(i)}(1)} = P^{(i)} = P_t/M$, $\mathcal{I} = \mathcal{M}$) yields sum DoF at most $n$, which is achieved by choosing $\mathbf{Q}^{(i)}_{\pi^{(i)}(1)}$ equal to a scalar matrix.

3. Applying superposition of multicast and unicast in $M' < M$ channels (i.e., $P^{(i)}_{\pi^{(i)}(1)} = \alpha_i P^{(i)}$ for $i \in \mathcal{I} \subset \mathcal{M}$) yields sum DoF $d_s = K(n - n'(M'/M)) + n'(M'/M)$, which is achieved by $\mathbf{Q}^{(i)}_0$ and $\mathbf{Q}^{(i)}_{\pi^{(i)}(1)}$ equal to scalar matrices. Note that $n < d_s < Kn$.

In the DoF sense, the superposition of multicast and unicasts smoothly bridges the two extremes between the maximum DoF $Kn$ with the multicast-only and the minimum DoF $n$ with the unicast-only. As opposed to sending multicast-only, increasing the number of subchannels with the superposition (i.e., increasing $M'$) lowers the overall sum DoF towards $n$. Consequently, a greater price must be paid by the multicast in order to accommodate the unicast in the high SNR regime. Clearly, the multicast needs to be favored in the high SNR region. Considering that in the presence of full CSIT, the sum DoF is $\min\{n_T, Kn_R\}$ [12], lack of CSI clearly limits the DoF for the unicast, similar to the observation made for unicast-only scenario [44]. The multicast, on the other hand, can achieve the full DoF even without CSI.

## IV. Power Allocation Problem

We now consider the power allocation problem between the subchannels and between the multicast and the single unicast messages. Throughout, we assume $\mu > 1$ since if $\mu \leq 1$, sending only the unicast messages is WSR optimal and the resulting power allocation between the unicast traffics can be obtained via an instance of the algorithms described below. It is worth noting that the aim of our proposed power allocation algorithms is to efficiently compute the highest achievable rates proposed by theorem 1 and reveal the benefits of employing the superposition transmission when both multicast and unicast messages are present. The maximum rates computed using these algorithms can provide useful performance benchmarks and guidelines for practical system designs. A practical low-complexity power allocation algorithm for system designs is beyond the scope of this work and is left for future study.

### A. Overview of the optimal power allocation algorithm

Using the results from Theorem 1, the rates for the multicast and unicast messages in subchannel $i$ are respectively simplified as

$$R^{(i)}_{0,k} = \mathbb{E}\left[\log\det\left(\mathbf{I} + \left(\mathbf{I} + \sigma^2_{i,k}\mathbf{H}\mathbf{Q}^{(i)}_{\pi^{(i)}(1)}\mathbf{H}^\dagger\right)^{-1} \sigma^2_{i,k}\mathbf{H}\mathbf{Q}^{(i)}_0\mathbf{H}^\dagger\right)\right]$$
$$= \mathbb{E}\left[\log\det\left(\mathbf{I} + \sigma^2_{i,k}\mathbf{H}\left(\mathbf{Q}^{(i)}_0 + \mathbf{Q}^{(i)}_{\pi^{(i)}(1)}\right)\mathbf{H}^\dagger\right)\right]$$
$$- \mathbb{E}\left[\log\det\left(\mathbf{I} + \sigma^2_{i,k}\mathbf{H}\mathbf{Q}^{(i)}_{\pi^{(i)}(1)}\mathbf{H}^\dagger\right)\right], \quad (18)$$

$$R^{(i)}_{\pi^{(i)}(1)} = \mathbb{E}\left[\log\det\left(\mathbf{I} + \sigma^2_{i,\pi^{(i)}(1)}\mathbf{H}\mathbf{Q}^{(i)}_{\pi^{(i)}(1)}\mathbf{H}^\dagger\right)\right], \quad (19)$$

where (18) is obtained by the log-determinant identity (11). Our problem (9) consists of maximization of the WSR $\sum_{i \in \mathcal{M}} \eta_i \sum_{k \in \mathcal{K}} \left(\mu_k R^{(i)}_{0,k} + R^{(i)}_{\pi^{(i)}(k)}\right)$ over $\{\mathbf{Q}^{(i)}_0(\boldsymbol{\mu}), \mathbf{Q}^{(i)}_{\pi^{(i)}(1)}(\boldsymbol{\mu})\}$ for all $i$, followed by minimization over $\boldsymbol{\mu}$. The global minimum in the minimization step can be efficiently found using the convex optimization methods. The maximization step is investigated next.

### B. Suboptimal power allocation by surrogate modeling

The considered maximization of the WSR in (9) is a nonconvex problem. Leveraging from Theorem 2, we search for the *local optimal solutions* $\mathbf{Q}^{(i)}_0 = \frac{1}{n_T}\left(P^{(i)} - P^{(i)}_1\right) \cdot \mathbf{I}$ and $\mathbf{Q}^{(i)}_{\pi^{(i)}(1)} = \frac{1}{n_T} P^{(i)}_1 \cdot \mathbf{I}$, where $P^{(i)}$ and $P^{(i)}_1$ are the total power allocated to subchannel $i$ and power allocated to the unicast message of the strongest user in subchannel $i$, respectively. This casts our WSR optimization problem into a power allocation problem.

We first define

$$\Phi(x) \triangleq \mathbb{E}_\mathbf{H}\left[\log\det\left(\mathbf{I} + \tfrac{x}{n_T}\mathbf{H}\mathbf{H}^\dagger\right)\right], \quad (20)$$

and an auxiliary function

$$\varphi(x) \triangleq \frac{\ln 2}{n_R} \frac{\partial \Phi(x)}{\partial x} = \frac{1}{n_R} \sum_{m=1}^{n_R} \mathbb{E}\left[\left(x + \tfrac{n_T}{d_m}\right)^{-1}\right], \quad (21)$$

which is proportional to the derivative of $\Phi(x)$ due to the equality $\log\det(\mathbf{I} + c\mathbf{H}\mathbf{H}^\dagger) = \sum_m \log(1 + cd_m)$ for any constant $c$, with $d_m$ being the $m^{\text{th}}$ largest eigenvalue of the Wishart matrix $\mathbf{HH}^\dagger$.[4] Then, a Lagrangian function of our problem is defined as, for some $\lambda \geq 0$,

$$\mathcal{L}\left(\boldsymbol{\mu}, \lambda, \left\{P^{(i)}, P_1^{(i)}\right\}_{\forall i}\right) \triangleq \sum_{i=1}^{M}\left[\eta_i\left(\sum_{k=1}^{K}\mu_k\left(\Phi\left(\sigma_{i,k}^2 P^{(i)}\right) - \Phi\left(\sigma_{i,k}^2 P_1^{(i)}\right)\right)\right.\right.$$
$$\left.\left.+\Phi\left(\sigma_{i,\pi^{(i)}(1)}^2 P_1^{(i)}\right)\right)\right] + \lambda\left(P_t - \sum_{i=1}^{M}P^{(i)}\right). \quad (22)$$

Power allocations over the subchannels, $P^{(i)}$ for all $i$, are first computed. For convenience, we define two *utility functions*, $u_0^{(i)}(x)$ and $u_1^{(i)}(x)$ for $x > 0$, as

$$u_0^{(i)}(x) \triangleq \sum_{k=1}^{K}\mu_k \sigma_{i,k}^2 n_R \varphi\left(\sigma_{i,k}^2 x\right) - \lambda \eta_i^{-1} \ln 2 \text{ and}$$
$$u_1^{(i)}(x) \triangleq \sigma_{i,\pi^{(i)}(1)}^2 n_R \varphi\left(\sigma_{i,\pi^{(i)}(1)}^2 x\right) - \lambda \eta_i^{-1} \ln 2. \quad (23)$$

Differentiating $\mathcal{L}$ with respect to $P^{(i)}$ yields

$$\eta_i^{-1} \ln 2 \frac{\partial \mathcal{L}}{\partial P^{(i)}} = \begin{cases} u_0^{(i)}\left(P^{(i)}\right), & \text{if } P_1^{(i)} < P^{(i)} \\ u_1^{(i)}\left(P^{(i)}\right), & \text{if } P_1^{(i)} = P^{(i)}, \end{cases} \quad (24)$$

and hence the optimal $P^{(i)}$'s are the zeros of the utility functions (illustrated in Fig. 5 for the case $P_1^{(i)} < P^{(i)}$). Since the utility functions are both monotonically decreasing and convex for $x > 0$ and the probability density function of $d_m$ are known (see e.g., [45], [46]), these zeros can be found by using standard root-finding methods, albeit computing the utility functions would involve numerical integrations.

The above procedure (and the subsequent procedures as described later) can be greatly simplified if we approximate the utility functions by their surrogate models. Specifically, we use the surrogate function defined by

$$\hat{\varphi}(x) \triangleq (1 + \alpha x)^{-1} \quad (25)$$

with parameter $\alpha$ that depends only on $n_T$ and $n_R$ to approximate the auxiliary function $\varphi(x)$. In Appendix C, it is shown that the surrogate function $\hat{\varphi}(x)$ closely matches $\varphi(x)$ over a wide range of $x > 0$ with the right choice of $\alpha$ (as a rule of thumb, $\alpha \approx 1$ in most scenarios of interest) and converges to $\varphi(x)$ as $n_T$ approaches infinity. The surrogate model allows us to approximate the optimal $P^{(i)}$ by the zeros of the following (differentiable and convex) utility functions:

$$\hat{u}_0^{(i)}(x) \triangleq \sum_{k=1}^{K}\mu_k \sigma_{i,k}^2 n_R \hat{\varphi}\left(\sigma_{i,k}^2 x\right) - \lambda \eta_i^{-1} \ln 2 \text{ and}$$
$$\hat{u}_1^{(i)}(x) \triangleq \sigma_{i,\pi^{(i)}(1)}^2 n_R \hat{\varphi}\left(\sigma_{i,\pi^{(i)}(1)}^2 x\right) - \lambda \eta_i^{-1} \ln 2. \quad (26)$$

---

[4] The expectations in $\Phi(x)$ and $\varphi(x)$ may be computed using a closed form expression (Eq. (4) in [45]), by numerical integration using eigenvalue distribution of the Wishart matrix (Eq. (42) in [45]), or via offline computation and a look-up table.

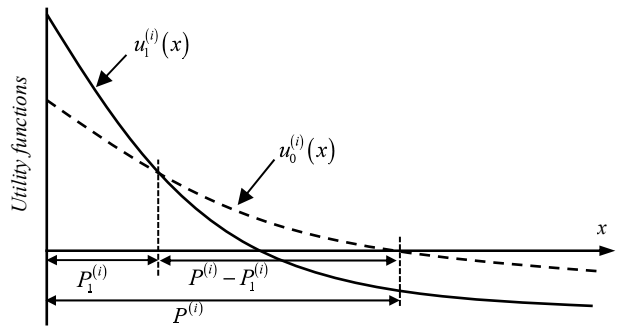

Fig. 5. An illustration of the utility functions and the corresponding optimal power allocations (24) and (29).

Let the *unique* zeros of $\hat{u}_0^{(i)}(x)$ and $\hat{u}_1^{(i)}(x)$ be denoted by $\hat{z}_0^{(i)}$ and $\hat{z}_1^{(i)}$, respectively. $\hat{z}_0^{(i)}$ may be quickly found using standard root finding methods such as the Newton-Raphson, while $\hat{z}_1^{(i)}$ is given in a closed form:

$$\hat{z}_1^{(i)} = \frac{1}{\alpha}\left(\frac{n_R \eta_i}{\lambda \ln 2} - \left(\sigma_{i,\pi^{(i)}(1)}^2\right)^{-1}\right). \quad (27)$$

In summary, the optimal power allocated to subchannel $i$ is approximately $\hat{z}_0^{(i)}$ if $P_1^{(i)} < P^{(i)}$, $\hat{z}_1^{(i)}$ if $P_1^{(i)} = P^{(i)}$, or a value 0 since power cannot be negative. Since $\mathcal{L}$ is an increasing function of $P^{(i)}$ until where its partial derivative is zero, it's clear that the largest possible $P^{(i)}$ must be chosen, i.e.,

$$P^{(i)} = \max\left\{\hat{z}_0^{(i)}, \hat{z}_1^{(i)}, 0\right\}. \quad (28)$$

The Lagrange multiplier $\lambda$, on the other hand, may be found by the standard waterfilling method. That is, the waterline $1/\lambda$ is incremented until the computed power in each subchannel using (28) adds to the total available power $P_t$.

Once $P^{(i)}$'s and $\lambda$ are computed, the optimal power split between the multicast and unicast messages are found. When $0 < P_1^{(i)} < P^{(i)}$, taking a partial derivative of $\mathcal{L}$ with respect to $P_1^{(i)}$ yields a difference of the utility functions:

$$\eta_i^{-1} \ln 2 \frac{\partial \mathcal{L}}{\partial P_1^{(i)}} = u_1^{(i)}\left(P_1^{(i)}\right) - u_0^{(i)}\left(P_1^{(i)}\right), \quad (29)$$

which implies that the optimal $P_1^{(i)}$ is where the two utility functions cross as illustrated in Fig. 5. This is a difference of convex (DC) functions with potentially arbitrary number of zeros. Fortunately, applying the surrogate approximation again yields a unique solution. Specifically, replacing the utility functions by their approximations, $\hat{u}_0^{(i)}(x)$ and $\hat{u}_1^{(i)}(x)$ from (26), and solving for $\partial \mathcal{L}/\partial P_1^{(i)} = 0$, the solution $P_1^{(i)}$ (if it exists) is the zero of the following function:

$$\hat{g}^{(i)}(x) = -1 + \sum_{k=1}^{K}\mu_k \frac{\sigma_{i,\pi^{(i)}(1)}^{-2} + \alpha x}{\sigma_{i,k}^{-2} + \alpha x}. \quad (30)$$

Due to $\sigma^2_{i,\pi^{(i)}(1)} > \sigma^2_{i,k}$ for all $k \neq \pi^{(i)}(1)$, $\hat{g}^{(i)}(x)$ is continuous and monotonically increasing in $x > 0$ with the asymptote $\mu - 1$ that is strictly positive for $\mu > 1$. Therefore, $\hat{g}^{(i)}(x)$ can cross the value zero exactly once if it initially starts negative at $x = 0$. Denoting the unique zero of $\hat{g}^{(i)}(x)$ by $\hat{z}^{(i)}$, the optimal $P_1^{(i)}$ is approximately $\hat{z}^{(i)}$ which is easily computed using the Newton-Raphson method. On the other hand, if $\hat{g}^{(i)}(0) \geq 0$, $\hat{g}^{(i)}(x)$ never crosses zero in $x > 0$, and it is optimal to simply set $P_1^{(i)} = 0$. It is worth noting that the condition $\hat{g}^{(i)}(0) \geq 0$ is equivalent to $\sum_{k \in \mathcal{K}} \mu_k \sigma^2_{i,k} \geq \sigma^2_{i,\pi^{(i)}(1)}$, and the resulting optimal transmission policy matches with the condition 4.1 of Theorem 1.

In summary, if $\sum_{k \in \mathcal{K}} \mu_k \sigma^2_{i,k} \geq \sigma^2_{i,\pi^{(i)}(1)}$, then it is optimal to set $P_1^{(i)} = 0$, else $P_1^{(i)} = \min\{\hat{z}^{(i)}, P^{(i)}\}$. The overall power allocation algorithm using the surrogate modeling is given in full detail below.

---

**Algorithm 1:** Proposed Resource Allocation over OFDMA Resources using Surrogate Modeling

**STEP 0: Initialization**
  Initialize $\boldsymbol{\mu} = [\mu_1, \cdots, \mu_K]$ so that $\mu_k \geq 0$ for all $k$ and $\sum_{k \in \mathcal{K}} \mu_k = \mu \ (>0)$.

**STEP 1.1: Determine the power split between subchannels ($P^{(i)}$ for all $i$)**
  Set an initial waterline $1/\lambda \ (>0)$ and a tolerance level $tol \ (\ll 1)$.
  
  While $\left| \sum_{i=1}^M P^{(i)} - P_t \right| > tol$, for all $i = 1, 2, \ldots, M$,
    Adjust waterline $1/\lambda$ (increment if $\sum_i P^{(i)} < P_t$; decrement otherwise.)
    Compute $\hat{z}_0^{(i)}$, i.e., the zero of $\hat{u}_0^{(i)}(x)$ from (26), using the Newton-Raphson algorithm.
    Compute $\hat{z}_1^{(i)}$ using (27).
    Set $P^{(i)} = \max\{\hat{z}_0^{(i)}, \hat{z}_1^{(i)}, 0\}$.
  End while

**STEP 1.2: Determine the power split between the multicast and unicast**
  For $i = 1, 2, \ldots, M$,
    If $P^{(i)} = 0$, set $P_{\pi^{(i)}(1)}^{(i)} = 0$. **(No power allocated in subchannel $i$)**
    Else if $P^{(i)} = \hat{z}_1^{(i)}$, set $P_1^{(i)} = P^{(i)}$. **(Unicast only)**
    Else if $\sum_{k=1}^{K} \mu_k \sigma^2_{i,k} \geq \sigma^2_{i,\pi^{(i)}(1)}$, set $P_1^{(i)} = 0$. **(Multicast only)**
    Else,
      Compute $\hat{z}^{(i)}$, i.e., the zero of (30), using the Newton-Raphson.
      Set $P_1^{(i)} = \min\{\hat{z}^{(i)}, P^{(i)}\}$. **(Superposition)**
    End if
    Set $P_0^{(i)} = P^{(i)} - P_1^{(i)}$.
  End For

**STEP 2: Compute the Lagrangian $\mathcal{L}$ using (22).**

**STEP 3: Convex Minimization**
  Using convex minimization methods, repeat steps 1 and 2 to search for $\boldsymbol{\mu}$ that minimizes $\mathcal{L}$. Return $P_1^{(i)}$ and $P_0^{(i)} \ \forall i$ at the minimizer.

---

### C. Further simplification using uniform power allocation over subchannels

Algorithm 1 can be further simplified with some performance penalty if uniform power allocation is assumed over the subchannels (i.e., by setting $P^{(i)} = P_t/M$) and only the power splits between the multicast and the unicast are optimized. In addition, $\hat{g}^{(i)}(x)$ from (30) may be lower bounded using Jensen's inequality as follows:

$$\hat{g}^{(i)}(x) \geq -1 + \mu \frac{\sigma^{-2}_{i,\pi^{(i)}(1)} + \alpha x}{\sum_{k \in \mathcal{K}} (\mu_k/\mu) \sigma^{-2}_{i,k} + \alpha x}. \quad (31)$$

Solving for the (unique) zero of this lower-bound yields an approximate $\hat{z}^{(i)}$ in a closed-form:

$$\hat{z}^{(i)} \approx \frac{1}{\alpha(\mu-1)} \left( \sum_{k=1}^{K} \frac{\mu_k}{\mu} \sigma^{-2}_{i,k} - \mu \sigma^{-2}_{i,\pi^{(i)}(1)} \right). \quad (32)$$

Applying these approximations yields a simplified resource allocation algorithm as shown in Algorithm 2. Note that Algorithm 2 replaces steps 1.1 and 1.2 of Algorithm 1 with closed form expressions.

---

**Algorithm 2:** Simplified Resource Allocation with Uniform Power Allocation over Subchannels

**Replace STEP 1.1 and 1.2 of Algorithm 1 with the following:**
  Set $P^{(i)} = P_t/M$ for all $i$.
  Set $P_1^{(i)} = \left[ \min\left\{ \frac{1}{\alpha(\mu-1)} \left( \sum_{k=1}^{K} \frac{\mu_k}{\mu} \sigma^{-2}_{i,k} - \mu \sigma^{-2}_{i,\pi^{(i)}(1)} \right), \frac{P_t}{M} \right\} \right]^+$ for all $i$.
  Set $P_0^{(i)} = (P_t/M) - P_1^{(i)}$ for all $i$.

---

## V. SYSTEM LEVEL SIMULATION RESULTS

System level simulation results of the proposed multicast and unicast superposition transmission are now presented. Table 1 lists the detailed simulation parameters, which generally follow the rural macrocell (RMa) scenario from 3GPP TR 38.901 [47] whose characteristics include continuous wide area coverage supporting high speed vehicles. The users are dropped randomly and uniformly over a hexagonal 19-site network with 3 sectors per site, and the results are provided for the users attached to the north facing sector of the central site. Once the channel SNRs $\sigma_{i,k}^2$ of the individual users are obtained from the system level simulations, the Rayleigh MIMO channels $\mathbf{H}_k^{(i)}$ for all $i$ and $k$ are independently generated from distribution $\mathcal{CN}(0, \sigma_{i,k}^2 \eta_i)$. We consider OFDMA systems with $M = 10$ subchannels with equal number of REs in each subchannel (i.e., $\eta_i = 1/M = 0.1$), and transmitters and users with $n_T = 8$ and $n_R = \{2, 4\}$ antennas, respectively. All results are averaged over 10,000 independent and uniform user drops.

Table I. System level simulation parameters based on RMa 3GPP TR 38.901

| | |
|---|---|
| **Carrier frequency** | 700 MHz |
| **Simulation bandwidth** | 10 MHz |
| **Intersite distance (ISD)** | 1732 m |
| **Cell layout** | Hexagonal, 19 macro sites, 3 sectors/site |
| **Sector antenna radiation patterns** | Based on Table 7.3-1 in [47] |
| **Transmit power per sector** | 46 dBm |
| **Heights** | Base-station (BS) 35 m, User-equipment (UE) 1.5 m |
| **BS & UE antenna gains** | BS 8 dBi, UE 0 dBi |
| **UE noise figure** | 7 dB |
| **UE deployment** | 50% indoor, 50% in car |
| **Pathloss model** | Based on section 7.4 in [47] |
| **Shadowing** | Log-normal fading with 7 dB standard dev.; Exponential geographic correlations [48] with correlation distance $d_{corr}$ = 120 m |
| **UE attachment** | Attached to the sector with strongest wideband SINR |

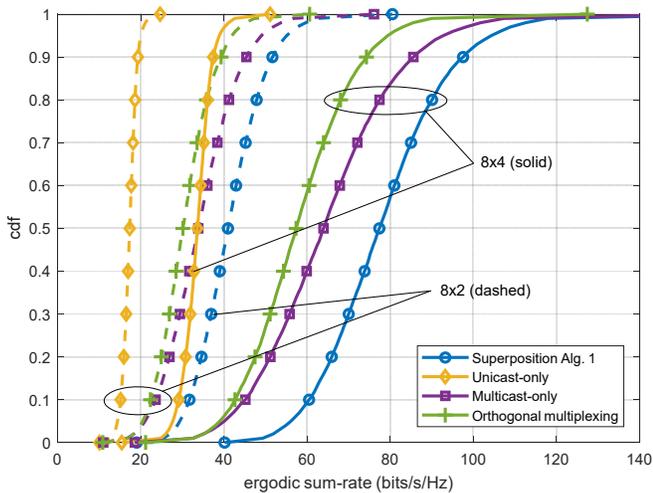

Fig. 6. CDF of average sum-rates (bits/s/Hz) of various resource allocation schemes from the system level simulations with 30 users with 8x2 or 8x4 MIMO settings. Superposition Alg. 1 is the multicast and unicast superposition transmission. All resource allocation schemes are optimized to maximize their respective ergodic sum-rates.

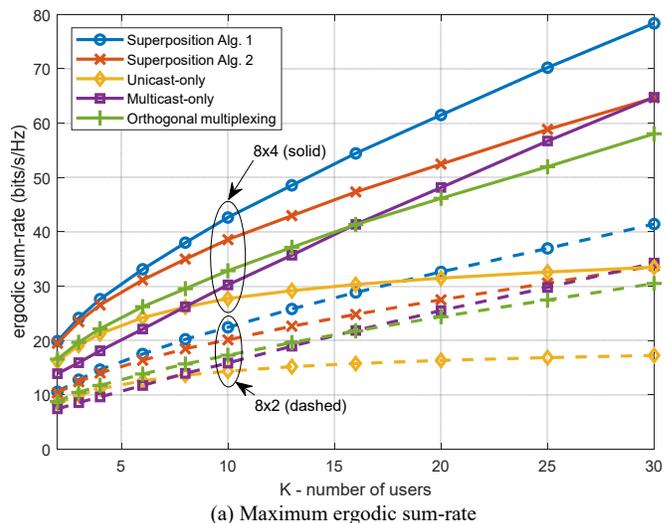

(a) Maximum ergodic sum-rate

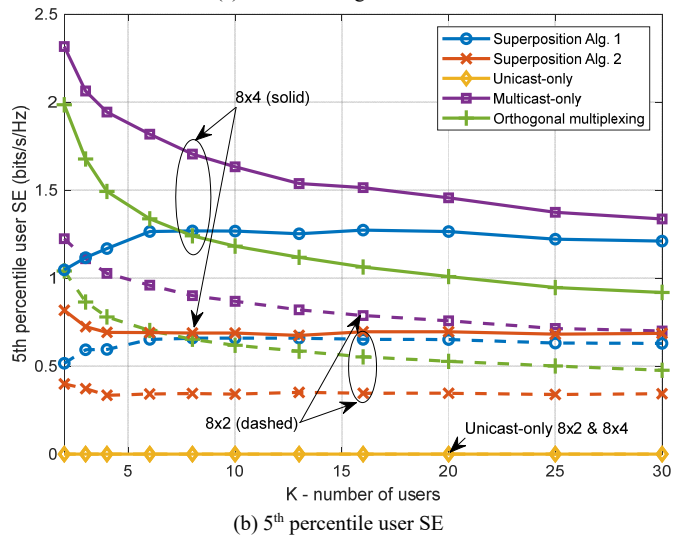

(b) 5th percentile user SE

Fig. 7. Average sum-rates and 5th percentile user SE versus number of users of various resource allocation schemes from the system level simulations. Superposition Alg. 1 and 2 are the multicast and unicast superposition transmissions. All resource allocation schemes except for Alg. 2 are optimized to maximize their respective ergodic sum-rates. Alg. 2 uses uniform power allocation over the subchannels.

Fig. 6 shows cumulative distribution functions (CDFs) of the average sum-rates $KR_0 + \sum_{k \in \mathcal{K}} R_k$ for $K = 30$ users. Fig. 7 shows the average sum-rates and the 5th percentile user spectral efficiency (i.e., the 5th percentile point of the CDFs, estimated from all possible user locations [49]) versus number of users. The proposed multicast and unicast superposition transmission (Algorithms 1 and 2) as well as competing schemes including the orthogonal multiplexing (OM), the unicast-only (UO), and the multicast-only (MO) transmissions are simulated. All resource allocation schemes except for Algorithm 2 are optimized to maximize their respective ergodic sum-rates. In OM, 50% of the subchannels are assumed to be allocated to multicast and the remaining 50% are allocated to unicast. The subchannel allocations for multicast and unicast and the power distributions over subchannels are optimized to maximize their ergodic sum-rates. In UO, all subchannels are assigned to unicast and this models a network that does not implement the multicast feature. In MO, all subchannels are assigned to multicast. Optimal covariances in UO and MO are known to be scalar matrices and hence the optimal resource allocation algorithms can be obtained from Algorithm 1 by setting $\mu = 0$ for UO or $P_1^{(i)} = 0$, $\forall i$ for MO. All algorithms use the surrogate approximations to simplify the optimizations.

To test the accuracy of the surrogate approximations, an optimization algorithm that directly solves the original ergodic WSR maximization problem (6) (without using the surrogate approximation) has also been implemented by using the MATLAB built-in solver *fmincon*. The computed mean-squared-error between the sum-rates obtained from this direct solver and Algorithm 1 is observed to be as low as 0.019 with $K = 20$ users and 8×4 MIMO. Due to this negligible error, the results from the direct solver are omitted in the plots.

Results from Fig. 6 and Fig. 7 confirm that the resource allocation based on multicast and unicast superposition transmission (Algorithm 1) yields the highest sum-rates and provides more than 95%, 27%, and 33% gains over UO, MO, and OM, respectively, at 20 users for 8×4 MIMO – suggesting that a large performance gain can be realized by enabling joint multicast and unicast feature in realistic network environments when users demand a multicast content. Similar gains were observed for 8×2 MIMO. In addition, multicast and unicast superposition outperforms OM in terms of the average sum-rates while delivering consistent data rates to the cell-edge users regardless of the number of users. It is worth noting that superposition transmission beyond two signals cannot improve the sum-rates due to Theorem 1. As expected from the DoF analysis, average sum-rates for multicast and unicast are seen to grow by a factor of $K \min(n_T, n_R)$ and $\min(n_T, n_R)$, respectively, hence benefiting from the availability of multiple antennas. Algorithm 2 is shown to have some performance penalties compared to Algorithm 1 due to using uniform power allocation across the subchannels but remains competitive for $K<30$ even when it is compared with other schemes that are using optimal power allocations.

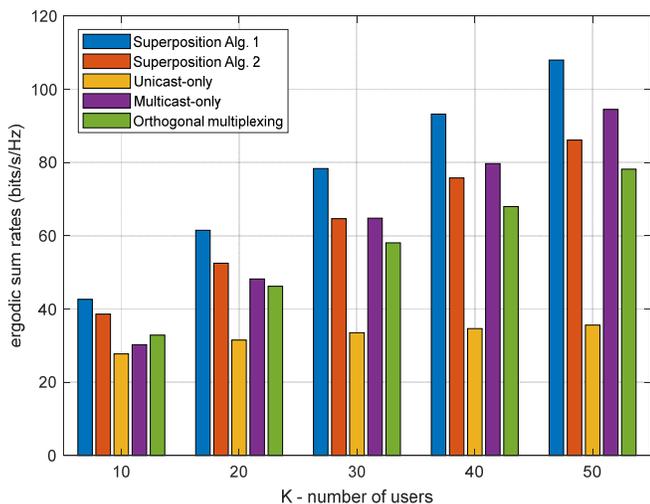

Fig. 8. Average sum-rates of various resource allocation schemes for up to 50 users in the system level simulations. All resource allocation schemes except for Alg. 2 are optimized to maximize their respective ergodic sum-rates. Alg. 2 uses uniform power allocation over the subchannels.

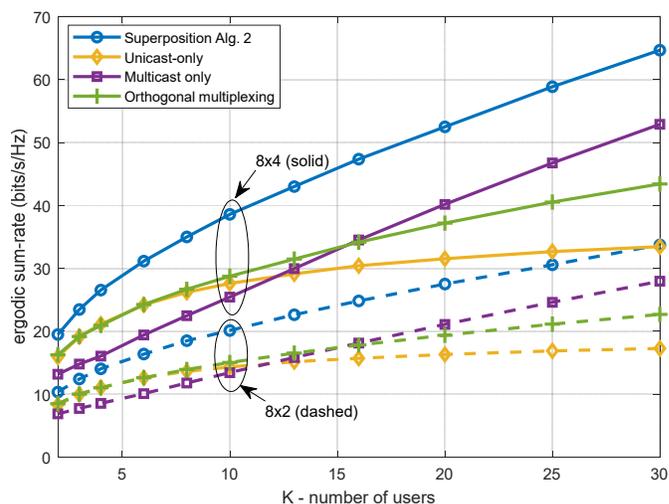

(a) Maximum ergodic sum-rate

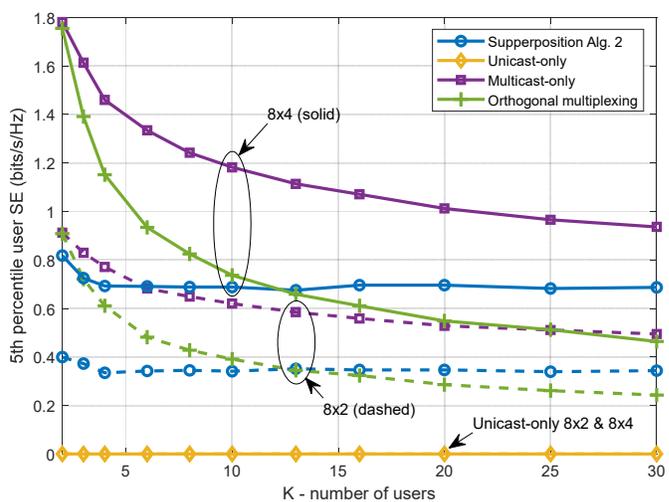

(b) 5$^{th}$ percentile user SE

Fig. 9. Average sum-rates and 5$^{th}$ percentile user SE versus number of users when the resource allocation schemes are using uniform power allocations over the subchannels. Superposition Alg. 2 is the multicast and unicast superposition transmission.

Fig. 8 extends the results from Fig. 7 to 50 users with 8x4 MIMO channels. We observe that the same trend continues into larger number of users where the multicast and unicast superposition transmission can provide 203%, 14%, and 38% gains over UO, MO, and OM, respectively, at 50 users.

Fig. 9 shows the performances of the various resource allocation schemes when power is uniformly distributed across the subchannels, which simplifies each algorithm albeit incurring some performance penalties. It can be observed multicast and unicast superposition transmission with uniform power allocation (Algorithm 2) still provides the highest sum-rates among all schemes (66%, 30%, and 41% gain over UO, MO, and OM, respectively, at $K = 20$ with 8×4 MIMO) and delivers consistent data rates to the cell-edge users. In addition, comparing with Fig. 7, the performance penalty for all resource allocation schemes from using uniform power allocation is negligible for small number of users but growing with the number of users (10.5% at $K = 10$ and 21% at $K = 30$).

Algorithms 1 and 2 for multicast and unicast superposition transmission dynamically select between unicast-only, multicast-only, and superposition transmission in each of the subchannels depending on the channel SNRs. Fig. 10 shows the proportion of the three modes being selected in the system level simulations. At 30 users, the superposition is selected for nearly 80% of the time, followed by unicast-only for about 20% of the time. Meanwhile multicast-only is rarely selected. Insights from Theorem 1 (also Fig. 4) indicate that the superposition is optimal when the ratio between channel SNRs of the strongest user and the weakest user exceeds the multicast rate reward $\mu$. When more users are dropped in a cell, the multicast rate reward $\mu = K$ (for the sum-rate) increases linearly, and so does the likelihood of greater channel variations between users. In the considered RMa scenario, Fig. 10 confirms that the ratio between channel qualities of the strongest and the weakest user increases as fast as the number of users, leading to superposition being selected most frequently. High percentage of superposition implies that there exists significant benefit to allow joint transmission of multicast and unicast in the simulated RMa channel environments.

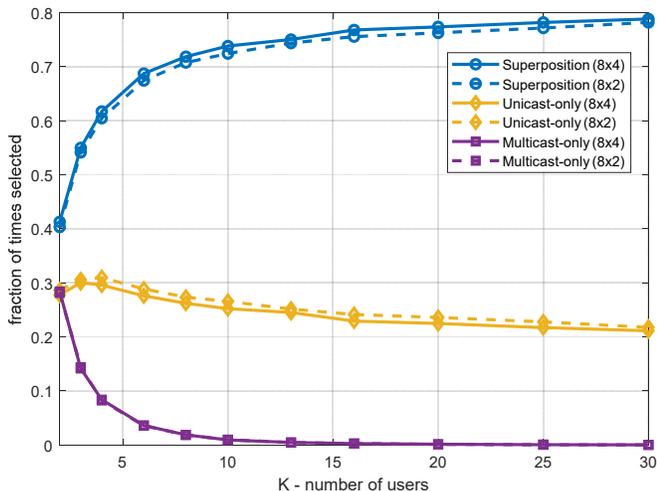

Fig. 10. Fraction of times the three transmission modes are being selected per subchannel by Algorithm 1 in the system level simulations.

## VI. Conclusion

In this paper, we demonstrate that multicast and unicast superposition transmission, with multi-layer MIMO in multi-user OFDMA systems with statistical CSIT, provides significant performance improvements over conventional orthogonal multiplexing, where the superposition with only two signals is shown to be sufficient to maximize the ergodic sum-rate. This approach is especially relevant for the delivery of common (e.g., popular live event) and independent (e.g., personalization) contents to a high number of users in deployments in the lower frequency bands operating in FDD mode, e.g., sub-1 GHz. We classify when it is optimal to send unicast only, multicast only, or their superposition in each subchannel, where the choice depends on the disparity between the best user and the worst user channel conditions and the minimum multicast rate target. The DoFs of multicast and unicast are shown to be $K \cdot \min(n_T, n_R)$ and $\min(n_T, n_R)$, respectively, and the superposition allows flexibly achieving a DoF between the two. Our developed resource allocation algorithms over OFDM resources and multicast/unicast messages show large performance gains due to the superposition of multicast and unicast messages in realistic network environments.

## Appendix A
## Proof of Theorem 2

By Theorem 1, without loss of generality we restrict our attention to non-negative diagonal matrices for both $\mathbf{Q}_0^{(i)}$ and $\mathbf{Q}_{\pi^{(i)}(1)}^{(i)}$, and consider the case $\sum_{k \in \mathcal{K}} \mu_k \sigma_{i,k}^2 < \sigma_{i,\pi^{(i)}(1)}^2$ (the only unknown case). As in the proof of theorem 1, we abbreviate $\mathbf{Q}_{\pi^{(i)}(1)}^{(i)}$ by $\mathbf{Q}_1^{(i)}$. We will need the following two lemmas.

*Lemma A.1:* Let $\Phi_k^{(i)}(\mathbf{Q})$ be defined as in (10) and $\phi_{j,k}^{(i)}(\mathbf{Q})$ be defined as

$$\phi_{j,k}^{(i)}(\mathbf{Q}) \triangleq \mathbf{e}_j^\dagger \mathbb{E}\left[\mathbf{H}^\dagger \left(I + \sigma_{i,k}^2 \mathbf{H}\mathbf{Q}\mathbf{H}^\dagger\right)^{-1} \mathbf{H}\right] \mathbf{e}_j, \quad (33)$$

where $\mathbf{e}_j$ is a unit vector with $j^{\text{th}}$ entry equal to 1. Suppose $\mathbf{Q}$ and $\mathbf{D}$ are two non-negative diagonal matrices and $P_j$ is the $j^{\text{th}}$ diagonal entry of $\mathbf{Q}$. Then,

$$\frac{\partial \Phi_k^{(i)}(\mathbf{Q}+\mathbf{D})}{\partial P_j} = \frac{\sigma_{i,k}^2}{\ln 2} \phi_{j,k}^{(i)}(\mathbf{Q}+\mathbf{D}).$$

*Proof:* Let $\mathbf{h}_j$ be $j^{\text{th}}$ column vector of $\mathbf{H}$. Using the matrix calculus (eq. (46) in [43]),

$$\frac{\partial \log\det\left(\mathbf{I}+\sigma^2 \mathbf{H}(\mathbf{Q}+\mathbf{D})\mathbf{H}^\dagger\right)}{\partial P_j}$$

$$= tr\left(\left(\mathbf{I}+\sigma^2 \mathbf{H}(\mathbf{Q}+\mathbf{D})\mathbf{H}^\dagger\right)^{-1} \frac{\partial \left(\sigma^2 \mathbf{H}\mathbf{Q}\mathbf{H}^\dagger\right)}{\partial P_j}\right)$$

$$= tr\left(\left(\mathbf{I}+\sigma^2 \mathbf{H}(\mathbf{Q}+\mathbf{D})\mathbf{H}^\dagger\right)^{-1} \sigma^2 \mathbf{h}_j \mathbf{h}_j^\dagger\right),$$

where the last equality is due to writing $\mathbf{H}\mathbf{Q}\mathbf{H}^\dagger = \sum_i P_i \mathbf{h}_i \mathbf{h}_i^\dagger$. The lemma is then proved by noting that $\mathbf{h}_j = \mathbf{H}\mathbf{e}_j$ and using the trace identity $tr(\mathbf{AB}) = tr(\mathbf{BA})$. ∎

*Lemma A.2:* Let $\phi_{j,k}^{(i)}(\mathbf{Q})$ be defined as in (33) for non-negative diagonal $\mathbf{Q}$ and let $P_{j_1}$ and $P_{j_2}$ be the $j_1^{\text{th}}$ and $j_2^{\text{th}}$ diagonal entries of $\mathbf{Q}$. Then, if $P_{j_1} = P_{j_2}$, $\phi_{j_1,k}^{(i)}(\mathbf{Q}) = \phi_{j_2,k}^{(i)}(\mathbf{Q})$.

*Proof:* We can find a permutation matrix $\Pi$ such that $\Pi \mathbf{e}_{j_2} = \mathbf{e}_{j_1}$ (which swaps $j_1$ and $j_2^{\text{th}}$ entries). Thus,

$$\phi_{j_1,k}^{(i)}(\mathbf{Q}) = \mathbf{e}_{j_2}^\dagger \mathbb{E}_\mathbf{H}\left[\Pi^\dagger \mathbf{H}^\dagger \left(I + \sigma_{i,k}^2 \mathbf{H}\mathbf{Q}\mathbf{H}^\dagger\right)^{-1} \mathbf{H}\Pi\right] \mathbf{e}_{j_2}$$
$$= \phi_{j_2,k}^{(i)}(\Pi^\dagger \mathbf{Q}\Pi),$$

where the last step follows from $\mathbf{H}$ being statistically invariant to any unitary transformations (i.e., $\mathbf{H} \sim \mathbf{H}\Pi$) and taking expectation with respect to $\mathbf{H}\Pi$. The proof is then complete by noting that $\Pi^\dagger \mathbf{Q}\Pi$ swaps $j_1$ and $j_2^{\text{th}}$ diagonal entries of $\mathbf{Q}$ and equals to $\mathbf{Q}$ if and only if $P_{j_1} = P_{j_2}$. ∎

*Remark 1:* Due to Lemma A.2, $\phi_{j,k}^{(i)}(\mathbf{Q})$ is equal for all $j$ when $\mathbf{Q}$ is a scalar matrix.

*Remark 2:* One can generalize Lemma A.2 to $\phi_{j_1,k}^{(i)}(\mathbf{Q}) \leq \phi_{j_2,k}^{(i)}(\mathbf{Q})$ if $P_{j_1} \geq P_{j_2}$, with equality if and only if $P_{j_1} = P_{j_2}$ using the concavity of (10) in $\mathbf{Q}$ and using Lemma A.1.

We now proceed to proving Theorem 2. Define a Lagrangian function of our maximization problem (13) by:

$$\mathcal{L} \triangleq \sum_{i=1}^{M} \eta_i \left( \sum_{k=1}^{K} \mu_k \Phi_k^{(i)}\left(\mathbf{Q}_0^{(i)}+\mathbf{Q}_1^{(i)}\right) - \sum_{k=1}^{K} \mu_k \Phi_k^{(i)}\left(\mathbf{Q}_1^{(i)}\right) + \Phi_{\pi^{(i)}(1)}^{(i)}\left(\mathbf{Q}_1^{(i)}\right) \right)$$
$$+ \lambda \left( P_t - \sum_{i=1}^{M} tr\left(\mathbf{Q}_0^{(i)}+\mathbf{Q}_1^{(i)}\right) \right), \tag{34}$$

for some $\lambda \geq 0$. Let $j^{\text{th}}$ diagonal entry of $\mathbf{Q}_0^{(i)}$ and $\mathbf{Q}_1^{(i)}$ be denoted by $P_{0,j}^{(i)}$ and $P_{1,j}^{(i)}$, respectively. Consider the following four cases:

**Case 1)** If $P_{0,j}^{(i)} = 0$ for all $j$ (i.e., $\mathbf{Q}_0^{(i)} = \mathbf{0}$), the objective function reduces to $\sum_i \eta_i \Phi_{\pi^{(i)}(1)}^{(i)}\left(\mathbf{Q}_1^{(i)}\right)$, which is trivially maximized by setting $\mathbf{Q}_1^{(i)} = \frac{P^{(i)}}{n_T}\mathbf{I}$.

**Case 2)** If $P_{0,j}^{(i)} > 0$ for some $j$, we need $\partial \mathcal{L}/\partial P_{0,j}^{(i)} = 0$ to satisfy the Karush-Kuhn-Tucker (KKT) conditions. Due to Lemma A.1, we have

$$\frac{\partial \mathcal{L}}{\partial P_{0,j}^{(i)}} = \frac{\eta_i}{\ln 2} \sum_{k=1}^{K} \mu_k \sigma_{i,k}^2 \phi_{j,k}^{(i)}\left(\mathbf{Q}_0^{(i)}+\mathbf{Q}_1^{(i)}\right) - \lambda,$$

and $\partial \mathcal{L}/\partial P_{0,j}^{(i)} = 0$ implies that we need

$$\sum_{k=1}^{K} \mu_k \sigma_{i,k}^2 \phi_{j,k}^{(i)}\left(\mathbf{Q}_0^{(i)}+\mathbf{Q}_1^{(i)}\right) = \eta_i^{-1} \lambda \ln 2$$

for all $j$ such that $P_{0,j}^{(i)} > 0$. In other words, for all $j_1 \neq j_2$, we need

$$\sum_{k=1}^{K} \mu_k \sigma_{i,k}^2 \phi_{j_1,k}^{(i)}\left(\mathbf{Q}_0^{(i)}+\mathbf{Q}_1^{(i)}\right) = \sum_{k=1}^{K} \mu_k \sigma_{i,k}^2 \phi_{j_2,k}^{(i)}\left(\mathbf{Q}_0^{(i)}+\mathbf{Q}_1^{(i)}\right).$$

This is satisfied if $\mathbf{Q}_0^{(i)} + \mathbf{Q}_1^{(i)}$ is a scalar matrix due to Lemma A.2.

**Case 3)** If $P_{1,j}^{(i)} = 0$ for all $j$ (i.e., $\mathbf{Q}_1^{(i)} = \mathbf{0}$), the optimal solution is trivially given by $\mathbf{Q}_0^{(i)} = \frac{P^{(i)}}{n_T}\mathbf{I}$.

**Case 4)** If $P_{1,j}^{(i)} > 0$ for some $j$, we need $\partial \mathcal{L}/\partial P_{1,j}^{(i)} = 0$ to satisfy the KKT conditions, which implies that

$$\sum_{k=1}^{K} \mu_k \sigma_{i,k}^2 \phi_{j,k}^{(i)}\left(\mathbf{Q}_0^{(i)}+\mathbf{Q}_1^{(i)}\right) - \sum_{k=1}^{K} \mu_k \sigma_{i,k}^2 \phi_{j,k}^{(i)}\left(\mathbf{Q}_1^{(i)}\right)$$
$$+ \sum_{k=1}^{K} \mu_k \sigma_{i,\pi^{(i)}(1)}^2 \phi_{j,\pi^{(i)}(1)}^{(i)}\left(\mathbf{Q}_1^{(i)}\right) = \eta_i^{-1} \lambda \ln 2$$

for all $j$ such that $P_{1,j}^{(i)} > 0$. The LHS of the above is equal for all $j$ if $\mathbf{Q}_0^{(i)}$ and $\mathbf{Q}_1^{(i)}$ are both scalar due to Lemma A.2. This completes the proof of Theorem 2. ∎

## APPENDIX B
## PROOF OF PROPOSITION 1

Let $\mathbf{Q} = \Delta \overline{\mathbf{Q}}$ with $\Delta = tr(\mathbf{Q})$ and $tr(\overline{\mathbf{Q}}) = 1$. For any $\alpha > 0$ and covariance matrix $\overline{\mathbf{Q}}$, we have

$$\mathbb{E}\left[\log \det\left(\mathbf{I} + \alpha \Delta \mathbf{H}\overline{\mathbf{Q}}\mathbf{H}^\dagger\right)\right] = \sum_{m=1}^{n'} \mathbb{E}\left[\log\left(1 + \alpha \Delta \lambda_m\right)\right]$$
$$\stackrel{\nearrow \Delta}{=} \sum_{m=1}^{n'} \mathbb{E}\left[\log\left(\alpha \Delta \lambda_m\right)\right] = n' \log \Delta + \sum_{m=1}^{n'} \mathbb{E}\left[\log\left(\alpha \lambda_m\right)\right], \tag{35}$$

where $\lambda_m$ is the $m^{\text{th}}$ largest eigenvalue of $\mathbf{H}\overline{\mathbf{Q}}\mathbf{H}^\dagger$ and $n'$ is the number of nonzero eigenvalues. Note that the rightmost term in (35) does not scale with $\Delta$. Using this observation, the ergodic sum-rate can be expressed in the limit $P_t \to \infty$ as

$$R_s(P_t) \stackrel{\nearrow P_t}{=} K \min_{k \in \mathcal{K}} \left\{ \sum_{i=1}^{M} \eta_i \sum_{m=1}^{n} \mathbb{E} \log\left(\sigma_{i,k}^2 P^{(i)} \lambda_m\right) \right.$$
$$\left. - \sum_{i=1}^{M} \eta_i \sum_{m=1}^{n'} \mathbb{E}\log\left(\sigma_{i,k}^2 P_{\pi^{(i)}(1)}^{(i)} \tilde{\lambda}_m\right) \right\} + \sum_{i=1}^{M} \eta_i \sum_{m=1}^{n'} \mathbb{E} \log\left(\sigma_{i,\pi^{(i)}(1)}^2 P_{\pi^{(i)}(1)}^{(i)} \tilde{\lambda}_m\right)$$
$$= K \sum_{i \in \mathcal{M}} \eta_i \left( n \log P^{(i)} - n' \log P_{\pi^{(i)}(1)}^{(i)} \right) + \sum_{i \in \mathcal{M}} \eta_i n' \log P_{\pi^{(i)}(1)}^{(i)} + o(1), \tag{36}$$

where $P^{(i)} = tr\left(\mathbf{Q}_0^{(i)} + \mathbf{Q}_{\pi^{(i)}(1)}^{(i)}\right)$, $\lambda_m$ and $\tilde{\lambda}_m$ are the $m^{\text{th}}$ largest eigenvalues of $\frac{1}{P^{(i)}} \mathbf{H}\left(\mathbf{Q}_0^{(i)} + \mathbf{Q}_{\pi^{(i)}(1)}^{(i)}\right)\mathbf{H}^\dagger$ and $\mathbf{H}\overline{\mathbf{Q}}_{\pi^{(i)}(1)}^{(i)}\mathbf{H}^\dagger$, respectively, and (36) is obtained by separating terms that do not scale with $P_t$ similar to (35). For $P_t$ large enough, the optimal power waterfilling across subchannels approaches equal power allocation $P^{(i)} = P_t/M$. Letting $P^{(i)} = P_t/M$ and collecting terms that do not scale with $P_t$ yields the desired expression. This completes the proof of Proposition 1. ∎

## APPENDIX C
## SURROGATE APPROXIMATION

The partial derivative of the ergodic WSR function with respect to the input power involves the following auxiliary function

$$\varphi(x) = \frac{1}{n_R} \sum_{m=1}^{n_R} \mathbb{E}\left[\left(x + \frac{n_T}{d_m}\right)^{-1}\right], \; x \geq 0$$

where $d_m$ is the $m^{\text{th}}$ largest eigenvalue of the Wishart matrix. While the function itself may be precomputed via a known probability distribution function of $d_m$ or Monte-Carlo methods, the optimization problem that we deal with needs to find the maximum of a difference of two $\varphi(x)$ which leads to a DC programming problem with a potentially arbitrary number of extrema. In this work, we approximate $\varphi(x)$ by a surrogate function defined as

$$\hat{\varphi}(x) = (1 + \alpha x)^{-1}, \; x \geq 0,$$

where the parameter $\alpha$ depends only on $n_T$ and $n_R$ and may be precomputed by data fitting methods. It's easy to see that both $\varphi$ and $\hat{\varphi}$ are continuous and monotonically decreasing towards zero, and both functions behave as $x^{-1}$ for $x \gg 1$ and $\alpha = 1$. Furthermore, with $\alpha = 1$ and as $n_T \to \infty$ while $n_R$ is fixed, $\mathbf{HH}^\dagger \sim n_T \mathbf{I}$ and $d_m \sim n_T$, and hence $\hat{\varphi}$ converges to the exact $\varphi$. It is also shown in section IV.B that with the surrogate function, the ergodic WSR optimization problem becomes concave with a unique maximum which can be computed easily by standard root-finding methods.

Table II. Parameter $\alpha(n_T, n_R)$ and corresponding mean-square-error (MSE)

| $\alpha(n_T,n_R)$ (MSE) | $n_R = 1$ | 2 | 4 | 8 | 16 |
|---|---|---|---|---|---|
| $n_T = 1$ | 1.306 (0.0058) | 2.284 (0.0021) | 4.267 ($5.39\times10^{-4}$) | 8.251 ($1.00\times10^{-4}$) | 16.212 ($1.56\times10^{-5}$) |
| 2 | 1.144 (0.0021) | 1.402 (0.0063) | 2.316 (0.0023) | 4.281 ($5.85\times10^{-4}$) | 8.253 ($1.08\times10^{-4}$) |
| 4 | 1.069 ($6.36\times10^{-4}$) | 1.160 (0.0024) | 1.435 (0.0058) | 2.324 (0.0024) | 4.281 ($5.60\times10^{-4}$) |
| 8 | 1.034 ($1.95\times10^{-4}$) | 1.071 ($6.46\times10^{-4}$) | 1.164 (0.0024) | 1.443 (0.0055) | 2.326 (0.0023) |
| 16 | 1.017 ($4.38\times10^{-5}$) | 1.034 ($1.78\times10^{-4}$) | 1.072 ($6.61\times10^{-4}$) | 1.165 (0.0024) | 1.445 (0.0055) |
| 32 | 1.008 ($1.24\times10^{-5}$) | 1.017 ($4.46\times10^{-5}$) | 1.034 ($1.73\times10^{-4}$) | 1.073 ($6.77\times10^{-4}$) | 1.166 (0.0024) |

Table 2 shows the parameter $\alpha$ that yields the smallest mean square error (MSE) over a finite range $x \in (0,100)$ for each $n_T$ and $n_R$ pair. These are obtained by numerical data fitting methods in MATLAB. Note that, as a rule of thumb, $\alpha$ is always close to 1 when $n_T \geq n_R$ and converges to 1 as $n_T$ increases. In the case of $n_T < n_R$, $\alpha$ is close to $n_R/n_T$.